\documentclass[11pt,a4paper]{article} 
\usepackage{jcappub}
\usepackage[normalem]{ulem}

\usepackage{enumitem}
\usepackage{geometry}
\newcommand\Tstrut{\rule{0pt}{2.9ex}} 

\usepackage{float}
\usepackage{epsfig}
\usepackage{epstopdf}
\usepackage{amsmath,dsfont}
\usepackage{hyperref}
\usepackage{amsfonts}
\usepackage{amssymb}
\usepackage{caption}
\usepackage{bm}
\usepackage{graphicx}
\usepackage{color}
\usepackage{verbatim}
\usepackage{subfigure}
\usepackage{blindtext, multicol}
\usepackage{setspace}
\usepackage{multirow}
\usepackage[capitalise]{cleveref}
\usepackage[activate={true,nocompatibility},final,kerning=true,factor=1100,stretch=10,shrink=10]{microtype}
\usepackage{academicons}
\usepackage[dvipsnames]{xcolor}
\usepackage{fontawesome5}

\definecolor{orcidlogocol}{rgb}{0.65, 0.807, 0.223}

\newcommand{\orcid}[1]{$\,$\href{https://orcid.org/#1}{\textcolor{orcidlogocol}{\faOrcid}}}

\usepackage[T1]{fontenc} 

\usepackage[sorting=none,style=numeric-comp]{biblatex}
\usepackage[english]{babel}
\usepackage{csquotes}
\DeclareUnicodeCharacter{0301}{\'{i}}
\usepackage[nottoc]{tocbibind}

\def\beq{\begin{equation}}
\def\eeq{\end{equation}}
\def\ber{\begin{eqnarray}}
\def\eer{\end{eqnarray}}
\def\benu{\begin{enumerate}}
\def\eenu{\end{enumerate}}

\def\l{\left}
\def\r{\right}

\newcommand{\sq}{\lower.25ex\hbox{\large$\Box$}}

\def\pa{\partial}
\def\f{\frac}
\def\mpl{m_{p}}

\def\ns{n_{_S}}
\def\nt{n_{_T}}

\def \lleq {\lower0.9ex\hbox{ $\buildrel < \over \sim$} ~}
\def \ggeq {\lower0.9ex\hbox{ $\buildrel > \over \sim$} ~}

\hypersetup{
	colorlinks,
	linkcolor={blue!70!black},
	citecolor={blue!70!black},
	urlcolor={blue!50!black}
}

\title{Canonical and Non-canonical Inflation in the light  of the recent BICEP/Keck results}
\author[a,b]{Swagat S. Mishra \orcid{0000-0003-4057-145X}}
\author[b]{and  Varun Sahni}

\affiliation[a]{School of Physics and Astronomy,  University of Nottingham, Nottingham, NG7 2RD, UK.}
\affiliation[b]{Inter-University Centre for Astronomy and Astrophysics,
Post Bag 4, Ganeshkhind, Pune 411~007, India}
\emailAdd{swagat.mishra@nottingham.ac.uk}
\emailAdd{varun@iucaa.in}

\date{\today}

\abstract{We discuss implications of the latest BICEP/Keck data release for inflationary models, with particular emphasis on  scalar fields with non-canonical Lagrangians  of the type ${\cal L} = X^\alpha - V(\phi)$. The observational upper bound on the tensor-to-scalar ratio, $r \leq 0.036$, implies that the whole family of monomial  power law potentials $V(\phi) \sim \phi^p$  are now ruled out  in the canonical framework  at $95\%$ confidence, which includes the simplest classic inflationary potentials such as  $\frac{1}{2}m^2 \phi^2$  and $\lambda \phi^4$.  Instead, current observations strongly favour asymptotically flat plateau  potentials. However, working in the non-canonical framework, we  demonstrate  that monomial  potentials, as well as the Higgs potential with its Standard Model self-coupling,   can easily be  accommodated by 
current CMB data.  We find striking  similarities  between the  $\lbrace n_{_S}, r\rbrace$ flow lines of  monomial potentials  in the non-canonical framework   and the   T-model $\alpha$-attractors in the canonical framework.
Significantly, $V(\phi)$ can originate from Planck scale initial values $V(\phi) \simeq m_p^4$ in
non-canonical models while in plateau-like canonical inflation the initial value of the potential
is strongly suppressed $V_{\rm plat}(\phi) \leq 10^{-10} m_p^4$.
This has bearing on the issue of initial conditions for inflation and allows for the equipartition of
the kinetic and potential terms in non-canonical models.}


\begin{document}
\maketitle

\section{Introduction}
The remarkable  progress  in Cosmology  over the past three decades,  reinforced by new theoretical insights and  fostered by a plethora of precision cosmological missions,  has resulted  in the so-called
   `standard scenario'  of the universe in the form of the spatially flat {\rm $\Lambda$}CDM model  \cite{Planck:2018vyg}. This model of `Concordance Cosmology' successfully describes the  evolution of our universe, both at the background as well as the  perturbative level, commencing  from almost a second after    the hot Big Bang 
until the present epoch \cite{Rubakov:2017xzr,Gorbunov:2011zzc}  (however, see \cite{Perivolaropoulos:2021jda}). 
Nevertheless, the success of this standard model relies upon several (seemingly) unnatural initial conditions 
which include, at the background level: the spatial homogeneity, isotropy  and near spatial flatness of
the universe, and at the perturbative level:  the existence of a  spectrum of  nearly scale-invariant adiabatic 
density fluctuations  (which seed structure formation in the universe).

        `Cosmic Inflation' has emerged as a leading prescription for describing the  very early universe prior to the commencement of the radiative hot Big Bang Phase  \cite{Starobinsky:1980te,Guth:1980zm,Linde:1981mu,Albrecht:1982wi,Linde:1983gd,Linde:1990flp,Baumann_TASI}. According to the 
inflationary paradigm, a transient epoch of at least 60-70 e-folds of  rapid accelerated expansion 
suffices  in  setting  natural initial conditions for cosmology in the
form of spatial flatness as well as  statistical homogeneity and isotropy on large angular scales  \cite{Guth:1980zm,Linde:1981mu,Albrecht:1982wi,Linde:1983gd,Linde:1990flp,Baumann_TASI}. 
Of equal significance is the fact that cosmic inflation  generates a spectrum of  initial scalar 
fluctuations  (via quantum fluctuations of  a scalar degree of freedom) which later seed  the formation of 
structure in the universe  \cite{Mukhanov:1981xt,Hawking:1982cz,Starobinsky:1982ee,Guth:1982ec}.  
In addition to  scalar perturbations, quantum fluctuations during inflation also create
 a spectrum of almost scale invariant tensor perturbations which later form gravity waves  \cite{Starobinsky:1979ty,Sahni:1990tx,Allen:1987bk}.

The simplest models of inflation comprising of a single scalar field, called the ‘inflaton’, which is minimally coupled to gravity, make several distinct predictions \cite{Tegmark:2004qd,Dodelson:2003ip}, most of which  have received spectacular observational confirmation, particularly from the latest  CMB missions  \cite{Planck_inflation}.   
Current observational data lead
to 
a scenario in which  the inflaton $\phi$ slowly rolls down a shallow potential $V(\phi)$ 
thereby giving rise to a 
quasi-de Sitter early stage of near-exponential expansion.

As mentioned earlier, both scalar and tensor perturbations are generated during inflation. 
The latter constitute the relic gravitational wave background (GW) which imprints
    a distinct signature on the  Cosmic Microwave Background (CMB) power spectrum in the form of the
 B-mode polarization \cite{Planck_inflation}. The amplitude of these relic GWs   provides us information about the inflationary energy scale while  their spectrum  enables us to access  general properties  of the epoch of reheating,  being exceedingly sensitive to the post-inflationary
equation of state \cite{Sahni:1990tx,Mishra:2021wkm}.    The amplitude of inflationary tensor fluctuations, 
relative to that of scalar fluctuations, is usually characterised by the tensor-to-scalar ratio $r$. Different models of inflation predict different values of $r$ which is sensitive to the gradient of the inflaton  potential $V'(\phi)=\frac{dV(\phi)}{d\phi}$ relative to its height $V(\phi)$. Convex potentials predict  large values for $r$, 
while concave potentials predict  relatively small values of $r$. While the spectrum of  inflationary tensor fluctuations has not yet been observed, current CMB observations are able to  place an upper bound on the tensor-to-scalar 
ratio on large angular scales. In particular, the latest CMB observations\footnote{ Note that BICEP stands for Background Imaging of Cosmic Extragalactic Polarization.} of BICEP/Keck \cite{BICEP:2021xfz}, combined with 
those of the PLANCK mission \cite{Planck_inflation}, place the strong upper bound $r\leq 0.036$ (at $95\%$ confidence). In this work we discuss the implications of this observational bound on single field  inflationary models, with particular emphasis on scalar fields with a non-canonical Lagrangian density. 

    This most recent upper bound on $r$,  combined with the $2\sigma$ bound on scalar spectral index $n_{_S}$, has important consequences for single field canonical inflation. In particular, given $r\leq 0.036$, all monotonically increasing convex potentials, including the whole family of monomial potentials $V(\phi)\propto \phi^p$, are completely ruled out in the canonical framework. Among these strongly disfavoured models are the simplest classic inflaton potentials $\f{1}{2}m^2\phi^2$ and $\lambda \phi^4$. Instead, the observational  upper bound on $r$ appears to favour asymptotically flat potentials possessing one or two plateau-like wings;  
see \cite{Kallosh:2021mnu}.    
    
Despite their excellent agreement with observations, plateau potentials may face some theoretical
shortcomings \cite{Ijjas:2013vea}. It is sometimes felt that, being a theory of the very early universe, inflation
should commence at the Planck scale \cite{Linde:1983gd,Ijjas:2013vea,Belinsky:1985zd,Belinsky:1987gy,Linde:2014nna} with $\rho_\phi \sim \mpl^4$. By contrast plateau potentials
are extremely flat and imply $V(\phi) \leq 10^{-10} \, \mpl^4$. Although a large value of the kinetic term
$V(\phi) \ll {\dot\phi}^2 \sim \mpl^4$ can offset this difficulty, it implies that the universe cannot
commence from equipartition initial conditions $V(\phi) \sim {\dot\phi}^2 \sim \mpl^4$.
Moreover the small value of the inflaton potential, $V(\phi) \leq 10^{-10} \, \mpl^4$, precludes the presence of a 
large curvature term, $|\rho_K| \sim \mpl^4$,
 at the commencement of inflation. Indeed, since $\rho_K = -3\,m_p^2/a^2 < 0$ in a closed
universe, an initial value $|\rho_K| \gg V(\phi)$ would imply that the universe would begin to contract 
prior to the onset of inflation in plateau potentials.

These issues can be successfully tackled in non-canonical models in which a small value of
$r$ can be accommodated by changes to the non-canonical kinetic term. Thus small values of 
$r\leq 0.036$ can easily arise for convex potentials having the form $V(\phi) \propto \phi^p$ and $V(\phi) \propto \phi^{-p}$ 
\cite{Unnikrishnan:2012zu,Unnikrishnan:2013vga}. In particular, we shall show that the Standard Model (SM) Higgs field, which 
cannot source CMB-consistent inflation in the canonical framework (owing to its large self-coupling), can 
easily source inflation in the non-canonical framework\footnote{Note that the SM Higgs can also source inflation if it is non-minimally coupled to gravity  \cite{Fakir:1990eg,Bezrukov:2007ep,Mishra:2018dtg}}.
Consequently equipartition initial conditions can easily be accomodated in non-canonical inflationary models
and in canonical models based on the Margarita potential.  Additionally, we discover striking similarities between the  $\lbrace n_{_S}, r\rbrace$ flow lines of the  T-model $\alpha$-attractors  in the canonical framework and  monomial potentials in the non-canonical framework, which is one of the central results of our analysis. We also discuss the differences in the predictions of $\lbrace n_{_S}, r\rbrace$  for these two classes of models.  Similarly we show that the inverse  power law  potential $V(\phi) \sim \phi^{-p}$, which leads to  power law inflation in the non-canonical framework,  also satisfies the latest CMB bounds.

Our paper is organised as follows:  
after a brief introduction of inflationary  scalar field dynamics in section \ref{sec:Inf_dyn}, we 
discuss inflation in the canonical framework in section \ref{sec:inf_can}
highlighting the  implications of the latest CMB observations on  
plateau potentials. 
Section \ref{sec:margarita} demonstrates that
some of the difficulties faced by plateau models can be circumvented 
in the Margarita family of potentials which interpolate between potentials which are convex at large
$|\phi|$ and concave at moderate $|\phi|$. 
Inflation in the non-canonical framework is discussed in
section \ref{sec:inf_noncan}. The concluding section \ref{sec:discussion} is dedicated to   a discussion highlighting the major inferences drawn from this work.

We work in the units $c,\hbar =1$. The reduced Planck mass 
is defined  to be $\mpl \equiv 1/\sqrt{8\pi G} = 2.43 \times 10^{18}~{\rm GeV}$. We assume the background
 universe to be homogeneous and isotropic with
the metric signature $(-,+,+,+)$

\section{Inflationary Dynamics}
\label{sec:Inf_dyn}

In the single field inflationary paradigm, inflation is sourced by a  scalar field, called the inflaton field, which is  minimally coupled to gravity and its dynamics is described by the action
\begin{equation}
S[\phi]=\int d^{4}x\, \sqrt{-g}\; {\cal L}(X,\phi) \, ,
\label{eqn: action}
\end{equation}
where the Lagrangian density ${\cal L}(\phi , X)$
is a function of the field $\phi$ and the
kinetic term
\begin{equation}
X=\frac{1}{2}\pa_{\mu}\phi\; \pa^{\mu}\phi \, .\label{eqn: X-phi}
\end{equation}
Varying the action in Eq.~(\ref{eqn: action}) \textit{w.r.t} $\phi$ results in the
 evolution equation of the inflaton field 
\begin{equation}
\frac{\pa {\cal L}}{\pa \phi} - \left(\frac{1}{\sqrt{-g}}\right)\pa_{\mu}\left(\sqrt{-g}\frac{\pa {\cal L}}{\pa \left(\pa_{\mu}\phi\right)}\right) = 0.\label{eqn: EOM1}
\end{equation}
While, the energy-momentum tensor
associated with the inflaton  field is given by
\begin{equation}
T^{\mu\nu}
=\left(\f{\pa{\cal L}}{\pa X}\right)\, \left(\pa^{\mu}\phi\; \pa^{\nu}\phi\right)
- g^{\mu\nu}\, {\cal L}~.\label{eqn: SET}
\end{equation}
In order to study the background dynamics during inflation, we specialize  to a a homogeneous scalar field in a  spatially flat FLRW universe, described by the line element
\begin{equation} 
{\rm d}s^2  =  -{\rm d}t^2 + a^{2}(t) \;  \left[{\rm d}x^2 + {\rm d}y^2 + {\rm d}z^2\right]  \, ,
\label{eqn: FRW}
\end{equation}
and the energy momentum tensor
\begin{equation}
T^{\mu}_{\;\:\;\nu} = \mathrm{diag}\left(-\rho_{_{\phi}}, p_{_{\phi}},  p_{_{\phi}},  p_{_{\phi}}\right),
\end{equation}
where the energy density $\rho_{_{\phi}}$, and pressure $p_{_{\phi}}$, of the homogeneous inflaton condensate  are given by
\begin{eqnarray}
\rho_{_{\phi}} &=& \left(\f{\pa {\cal L}}{\pa X}\right)\, (2\, X)- {\cal L}\label{eqn: rho-phi} \, ,\\
p_{_{\phi}} &=& {\cal L}\label{eqn: p-phi} \, .
\end{eqnarray}
The background kinetic term is given by $X = -\f{1}{2}\dot{\phi}^2$.
The evolution of the scale factor $a(t)$ is governed by the two Friedmann equations
\begin{eqnarray}
 \left(\frac{\dot{a}}{a}\right)^{2}  &\equiv& H^2 = \f{1}{3m_p^2} \, \rho_{_{\phi}},\label{eqn: Friedmann eqn1}\\
\frac{\ddot{a}}{a} &\equiv& \dot{H} + H^2 = -\f{1}{6m_p^2}\, \left(\rho_{_{\phi}} + 3\,p_{_{\phi}}\right),\label{eqn: Friedmann eqn2}
\end{eqnarray}
where $H \equiv \dot{a}/a$ is the Hubble parameter. The scalar field energy density and pressure satisfy the energy-momentum conservation equation
\begin{equation}
{\dot \rho_{_{\phi}}} = -3\, H \left(\rho_{_{\phi}} + p_{_{\phi}}\right) \, .
\label{eqn: conservation eqn}
\end{equation}
The aforementioned  discussion is applicable to  scalar fields with both canonical and non-canonical kinetic terms. 

\section{Inflation in the canonical framework}
\label{sec:inf_can}
For a canonical scalar field  $\phi$, with a suitable  potential $V(\phi)$,  the Lagrangian density in Eq.~(\ref{eqn: action}) becomes
\begin{equation}
{\cal L}(X,\phi) = - X - V(\phi)  \, ,
\label{eqn: Lagrangian}
\end{equation}
Substituting Eq.~(\ref{eqn: Lagrangian}) into
Eqs.~(\ref{eqn: rho-phi})~and~(\ref{eqn: p-phi}), we obtain
\begin{eqnarray}
\rho_{_{\phi}} &=& \frac{1}{2}{\dot\phi}^2 +\;  V(\phi) \, ,\nonumber\\
p_{_{\phi}} &=& \frac{1}{2}{\dot\phi}^2 -\; V(\phi) \, , ~~
\label{eqn: p-model}
\end{eqnarray}
and hence,
the two Friedmann Eqs.~(\ref{eqn: Friedmann eqn1})~and~(\ref{eqn: Friedmann eqn2}),  and the conservation Eq.~(\ref{eqn: conservation eqn}) become
\ber
H^2 = \frac{1}{3m_p^2} \, \rho_{\phi} &=& \frac{1}{3m_p^2} \left[\frac{1}{2}{\dot\phi}^2 +V(\phi)\right],
\label{eq:friedmann1}\\
\dot{H} = \frac{\ddot{a}}{a}-H^2 &=&  -\frac{1}{2m_p^2}\, \dot{\phi}^2,
\label{eq:friedmann2}\\
 {\ddot \phi}+ 3\, H {\dot \phi} + V'(\phi) &=& 0 ~.
\label{eq:phi_EOM}
\eer
Expansion of space during the inflationary evolution at an epoch with scale factor $a$  is usually described by  the   number of e-folds  before the end of inflation, defined as 
\beq
N_e(a)  = \ln \l( \frac{a_e}{a} \r) =\int_{t}^{t_e} H(t) \, {\rm d}t,
\label{eq:efolds}
\eeq
where $a_e$ denotes the scale factor at the  
end of inflation. Hence $N_e  = 0$ corresponds to the end of inflation. Typically a period of near-exponential inflation lasting for at least 60-70 e-folds is required in order to address the fine tuning initial conditions of the standard hot Big Bang phase. Throughout the paper, we denote $N_*$ as the number of e-folds before the end of inflation when the CMB pivot scale $k_*=(aH)_*=0.05~{\rm Mpc}^{-1}$ made its Hubble-exit. Usually, one considers  $N_* \in [50,\,60]$ depending upon the details of reheating history in the post-inflationary universe.
 
The slow-roll regime of inflation, facilitated by the presence of the Hubble-drag term in Eq.~(\ref{eq:phi_EOM}), is conveniently characterised by  two kinematic Hubble slow-roll parameters $\epsilon_H$, $\eta_H$, defined by  \cite{Lyth:2009zz,Baumann_TASI} 
\ber
\epsilon_H &=& -\frac{\dot{H}}{H^2}=\frac{1}{2m_p^2} \, \frac{\dot{\phi}^2}{H^2} \, ,
\label{eq:epsilon_H}\\
\eta_H &=& -\frac{\ddot{\phi}}{H\dot{\phi}}=\epsilon_H  + \frac{1}{2\epsilon_H} \, \frac{d\epsilon_H}{dN_e} \, ,
\label{eq:eta_H}
\eer
where the slow-roll conditions are defined as 
\beq
\epsilon_H,~\eta_H\ll 1 \, .
\label{eq:slow-roll_condition}
\eeq
The slow-roll regime  is also often characterised  by the dynamical potential slow-roll parameters  \cite{Baumann_TASI}, defined by
\begin{equation}
\epsilon_{_V}  = \frac{m_{p}^2}{2}\left (\frac{V'}{V}\right )^2 \, , ~~
\eta_{_V} = m_{p}^2 \, \left( \frac{V''}{V} \right) \, .
\label{eqn:slow_roll1}
\end{equation}
Under the slow-roll approximations, we have $\epsilon-H \simeq \epsilon_{_V}$ and $\eta_H \simeq \eta_{_V}-\epsilon_{_V}$. Since $H=\dot{a}/a$, using Eq.~(\ref{eq:epsilon_H}), we obtain
\beq
\f{\ddot{a}}{a} = \big( 1 - \epsilon_H \big) \, H^2 \, ,
\label{eq:inf_acc_cond}
\eeq
which demonstrates  that the universe accelerates, ${\ddot a} > 0$, when $\epsilon_{H} < 1$.
Using equation (\ref{eq:friedmann1}), the expression for  $\epsilon_{H}$ in (\ref{eq:epsilon_H}) reduces to $\epsilon_{H} \simeq \f{3}{2}\f{\dot{\phi}^2}{V}$  when
${\dot\phi}^2 \ll V$.  In fact, under the slow-roll conditions (\ref{eq:slow-roll_condition}), the Friedmann equations (\ref{eq:friedmann1}) and (\ref{eq:phi_EOM}) take the form
\ber
H^2 \simeq \f{1}{3m_p^2} V(\phi) \, , \\ \label{eq:friedmann_SR1}
\dot{\phi} \simeq - \f{V'(\phi)}{3H} \, .
\label{eq:friedmann_SR3}
\eer
Combining the  slow-roll Eqs.~(\ref{eq:friedmann_SR1}) and (\ref{eq:friedmann_SR3}) we  obtain the {\em slow-roll trajectory} described by 
\beq
 \dot{\phi} \simeq  - \f{m_p}{\sqrt{3}} \, \f{V_{,\phi}}{\sqrt{V(\phi)}} \, .
\label{eq:SR_trajectory}
\eeq
It is well known that the slow-roll trajectory is actually a local attractor for a variety of  different inflationary potentials, see   Refs.~\cite{Brandenberger:2016uzh, Mishra:2018dtg}.

\subsection*{{\bf Quantum fluctuations during inflation}}
In the single-field inflationary scenario, two gauge-invariant massless fields, one scalar and one transverse traceless tensor, get excited during inflation and receive quantum fluctuations that are  correlated over super-Hubble scales~\cite{Baumann:2018muz} at late times.   All cosmologically relevant fluctuations start out as being sub-Hubble, namely $k \gg aH$, at  early times. During inflation, as the comoving Hubble radius $\l( aH \r)^{-1}$ decreases with time, fluctuations eventually become super-Hubble, \textit{i.e} $k \ll aH$, where they remain frozen until their subsequent Hubble re-entry after the end of inflation.  The  gauge-invariant scalar degree of freedom $\cal R$ during inflation is  called the comoving curvature perturbation\footnote{Note that the comoving curvature perturbation $\cal R$, is also related to the curvature perturbation on uniform-density hypersurfaces, $\zeta$, and both are equal during slow-roll inflation as well as on super-Hubble scales, $k\ll aH$, in general (see  \cite{Baumann_TASI}).}   whose  dimensionless  primordial power-spectrum at super-Hubble scales   is given by \cite{Baumann:2018muz}
\beq
P_{\cal R} \equiv   \frac{k^3}{2\pi^2} \, |{\cal R}_k|^2 \Big|_{k\ll aH} ~,
\label{eq:MS_Ps}
\eeq
where ${\cal R}_k$ is the Fourier transformation of $\cal R$.   During slow-roll inflation, imposing  suitable Bunch-Davies initial conditions   \cite{Bunch:1978yq} in the sub-Hubble regime $k \gg aH$,  the subsequent computation of the super-Hubble  power spectrum of $\cal R$ leads to  the famous slow-roll approximation formula   \cite{Baumann_TASI}
\beq
P_{\cal R} = \frac{1}{8\pi^2}\left(\frac{H}{m_p}\right)^2\frac{1}{\epsilon_H}~.
\label{eq:Ps_slow-roll}
\eeq

On  large cosmological scales which are accessible to  CMB observations, the scalar  power spectrum typically takes the form of a  power law represented by 

\beq
P_{\cal R}(k)=A_{_S}\left(\frac{k}{k_*}\right)^{\ns-1},
\label{eq:Ps_power-law}
\eeq
where $A_{_S}=P_{\cal R}(k_*)$ is the amplitude of the scalar power spectrum at the pivot scale \footnote{Note that in general, $k$ may correspond to any observable CMB scale in the range $k\in \left[0.0005,0.5\right]~{\rm Mpc}^{-1}$. However, in order to derive constraints on the inflationary observables $\lbrace \ns,r \rbrace$, we mainly focus on the CMB pivot scale, namely    $k \equiv  k_*=0.05~{\rm Mpc}^{-1}$.} $k =  k_*$, 

\beq
A_{_S} = \frac{1}{8\pi^2}\left(\frac{H}{m_p}\right)^2\frac{1}{\epsilon_H}\bigg\vert_{k=k_*}~.
\label{eq:As_slow-roll}
\eeq

 The scalar spectral tilt $\ns$, in the slow-roll regime 
is given by \cite{Baumann_TASI}

\ber
\ns - 1 \equiv  \frac{d\, \mathrm{ln} P_{\cal R}}{d\, \mathrm{ln} k} = 2\eta_H-4\epsilon_H ~.
\label{eq:ns}
\eer

Similarly the tensor power spectrum, in the slow-roll limit,  is represented by
\beq
P_{T}(k) = A_{_T}\left(\f{k}{k_*}\right)^{\nt},\label{eq:Pt_power-law}
\eeq
with the amplitude of tensor power spectrum at the CMB pivot scale  is given by  \cite{Baumann_TASI,Baumann:2018muz}
\beq
A_{_T}\equiv P_T(k_*) = \f{2}{\pi^2}\left(\frac{H}{\mpl}\right)^2\bigg\vert_{k=k_*} ~,
\label{eq:At_slow-roll}
\eeq
and the tensor spectral index (with negligible running) is given by
\beq
\nt = -2 \, \epsilon_H~.
\label{eq:nt_slow-roll}
\eeq
The tensor-to-scalar ratio $r$ is defined by
\beq
r \equiv \f{A_{_T}}{A_{_S}} = 16\, \epsilon_H~,
\label{eq:r_slow-roll}
\eeq
yielding the \textit{single field consistency relation} 
\beq
r = -8\, \nt~.
\label{eq:consis_slow-roll}
\eeq

We therefore find that the
 the slow-roll parameters $\epsilon_H$ and $\eta_H$ play a key role in characterising  the power spectra of scalar and tensor fluctuations during inflation. In the next section, we discuss the implications of the latest CMB observations for  the slow-roll 
parameters and for  other relevant inflationary observables. In order to relate the CMB observables to the inflaton potential $V(\phi)$, we work with the potential slow-roll parameters\footnote{ While for slow-roll inflation, we obtain constraints on the potential slow-roll parameters $\epsilon_{_V}, \, \eta_{_V}$ from the CMB data, we also present constraints on the Hubble slow-roll parameters $\epsilon_H, \, \eta_H$  because they directly represent the evolution of the scale factor during inflation.} defined in  equation (\ref{eqn:slow_roll1}).

 \subsection{Implications of the latest CMB observations for canonical inflation}
 \label{sec:Inf_CMB_obs}
 
 Consider a canonical scalar field  minimally coupled to gravity and having the potential
\beq
V(\phi)=V_0 \, f\left(\f{\phi}{m_p}\right)~.
\label{eq:CMB_pot}
\eeq
The potential slow-roll parameters (\ref{eqn:slow_roll1}) are given by 
\ber
\epsilon_{_V} = \f{\mpl^2}{2}\left(\f{f'}{f}\right)^2~,\label{eq:CMB_eV} \\
\eta_{_V} = \mpl^2\left(\f{f''}{f}\right)~.\label{eq:CMB_etaV}
\eer
 In the  slow-roll limit $\epsilon_{_V}, \, \eta_{_V} \ll 1 $, the scalar  power spectrum is given by the expression  (\ref{eq:Ps_power-law})
with the amplitude of scalar power  at the CMB pivot scale $k\equiv k_*=0.05~{\rm Mpc}^{-1}$ expressed as  \cite{Baumann_TASI}
\beq
A_{_S}\equiv P_{\cal R}(k_*) \simeq \f{1}{24\pi^2}\f{V_0}{\mpl^4}\f{f\left(\phi_k\right)}{\epsilon_{_V}(\phi_k)}\bigg\vert_{k=k_*}~,
\label{eq:CMB_As}
\eeq
and the scalar spectral index (with negligible running) is given by
\beq
\ns \simeq 1 + 2 \, \eta_{_V}(\phi_*) - 6 \, \epsilon_{_V}(\phi_*) ~,
\label{eq:CMB_ns}
\eeq
where $\phi_*$ is the value of the inflaton field at the Hubble exit of CMB pivot scale $k_*$.  
Similarly the amplitude of tensor power spectrum at the CMB pivot scale  is  given by
\beq
A_{_T}\equiv P_T(k_*) = \f{2}{\pi^2}\left(\frac{H_k^{\rm inf}}{\mpl}\right)^2\bigg\vert_{k=k_*} \simeq  \f{2}{3\pi^2}\f{V_0}{\mpl^4}f\left(\phi_k\right)\bigg\vert_{k=k_*}~,
\label{eq:CMB_At}
\eeq
where  $H_k^{\rm inf}$ is the  Hubble parameter during the Hubble exit of mode k. The tensor spectral index (\ref{eq:nt_slow-roll})  becomes 
\beq
\nt \simeq -2 \, \epsilon_{_V}(\phi_*)~,
\label{eq:CMB_nt}
\eeq
and the tensor-to-scalar ratio (\ref{eq:r_slow-roll}) can be written as  
\beq
r \simeq 16\, \epsilon_{_V}(\phi_*)~,
\label{eq:CMB_r}
\eeq
satisfying the single field consistency relation  (\ref{eq:consis_slow-roll}).
From the CMB observations of Planck 2018 \cite{Planck_inflation}, we have  
\beq
A_{_S} = 2.1\times 10^{-9}~,
\label{eq:CMB_As_obs}
\eeq
while the  $2\sigma$ constraint on the scalar spectral index is given by
\beq
n_{_S} \in [0.957,0.976]~.
\label{eq:CMB_ns_obs}
\eeq
Similarly the constraint on the tensor-to-scalar ratio $r$, from the  latest combined observations of Planck 2018  \cite{Planck_inflation} and  BICEP/Keck \cite{BICEP:2021xfz}, is given by
\beq
r \leq 0.036~,
\label{eq:CMB_r_obs}
\eeq
which translates into $A_{_T}\leq 3.6\times 10^{-2}\, A_{_S}$. Equation (\ref{eq:CMB_At}) helps place the following
 upper bound on the inflationary Hubble scale $H_k^{\rm inf}$ and the energy scale during inflation $E_{\inf}$
\ber
H_k^{\rm inf} \leq  4.7\times 10^{13}~{\rm GeV}~,\label{eq:CMB_H_k} \\
E_{\inf}\equiv \left[\sqrt{3} \,  m_p \, H_k^{\rm inf}\right]^{1/2} \leq 1.4\times 10^{16}~{\rm GeV}~~.\label{eq:CMB_T_inf}
\eer
Similarly the CMB bound on $r$ when combined with (\ref{eq:CMB_r}) translates into an
  upper bound on the first slow-roll parameter 
 \beq
\epsilon_H \simeq \epsilon_{_V} \leq 0.00225~,
\label{eq:CMB_eV_obs}
\eeq
 rendering the  tensor tilt from equation (\ref{eq:CMB_nt}) to be negligibly small
\beq
| \nt| \leq 0.0045~.
\label{eq:CMB_nt_obs}
\eeq
Given the upper limit on $\epsilon_{_V}$, using the CMB bound on $\ns$ from (\ref{eq:CMB_ns_obs}) in (\ref{eq:CMB_ns}), we infer that the  \textit{second slow-roll parameter is negative} and  obtain interesting upper and lower limits on its magnitude, given by
\beq
|\eta_H| \in [0.0075,0.0215]~.
\label{eq:CMB_etaV_obs}
\eeq
 The EOS $w_\phi$ of  the inflaton field is given by  
\beq
w_\phi = \f{\f{1}{2}\dot{\phi}^2-V(\phi)}{\f{1}{2}\dot{\phi}^2+V(\phi)} \simeq -1+\f{2}{3}\epsilon_{_V}(\phi)\, ,
\label{eq:CMB_w_obs}
\eeq
Therefore one finds from (\ref{eq:CMB_eV_obs}) the following constraint on the inflationary EOS at the pivot scale 
\beq
w_\phi \leq -0.9985 \, ,
\label{eq:CMB_w_inf_obs}
\eeq
implying that the expansion of the universe during inflation was near exponential (quasi-de Sitter like).

\begin{figure}[htb]
\begin{center}
\includegraphics[width=0.8\textwidth]{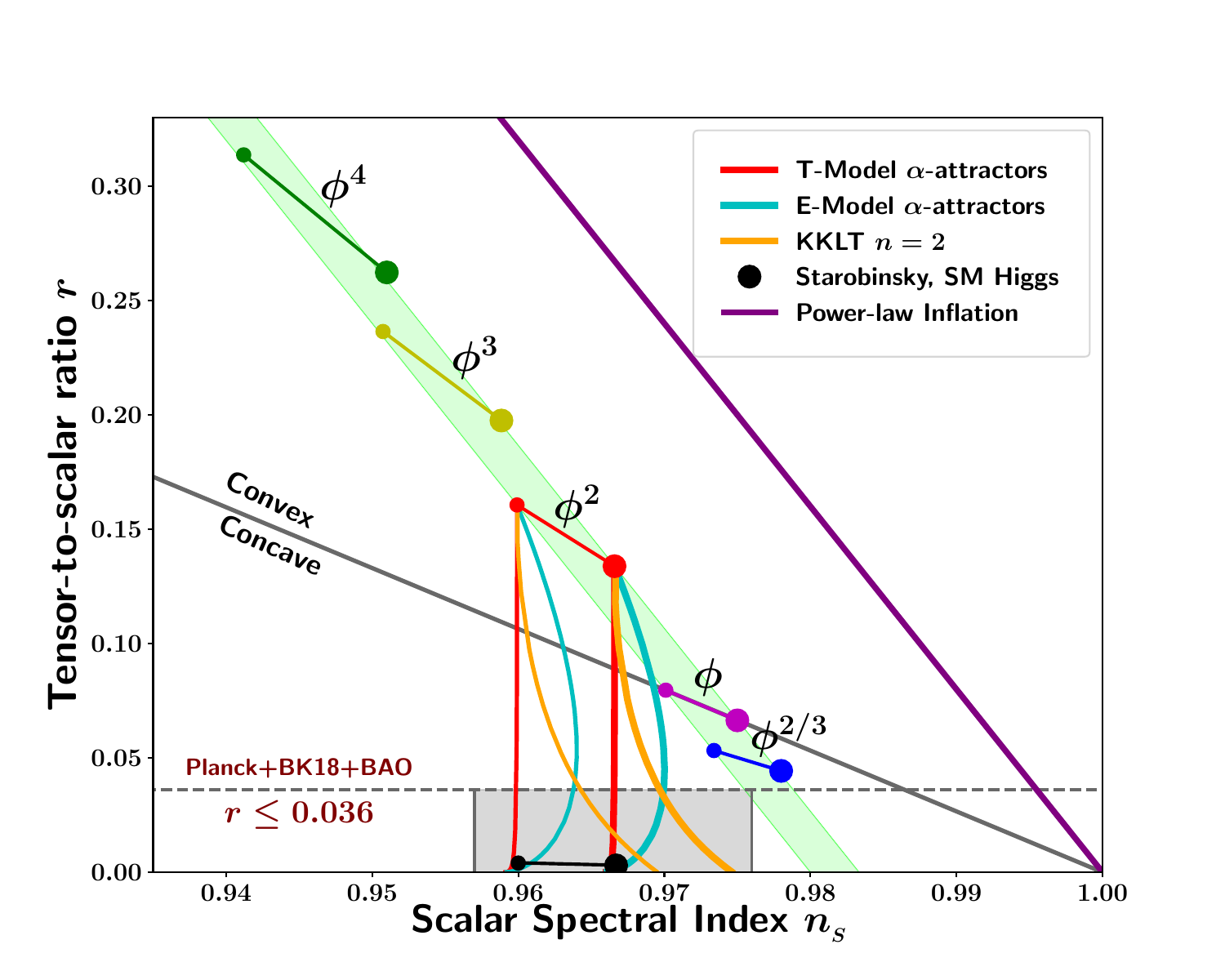}
\caption{This figure plots the  tensor-to-scalar ratio $r$ versus  the  scalar spectral index $\ns$  for a number of 
inflationary  potentials (the thinner and thicker dots and  curves correspond to $N_* = 50,\, 60$ respectively). 
These  include  predictions of  plateau potentials such as the Starobinsky model, the Standard Model Higgs inflation, the  T-model and E-model $\alpha$-attractors as well as the D-brane  KKLT inflation. The CMB 2$\sigma$ bound 
$0.957 \leq \ns \leq 0.976$
 and
 the upper bound on the tensor-to-scalar ratio  $r\leq 0.036$ are
 indicated by the shaded grey colour region. Given the upper bound on $r$, it is easy to see that observations favour concave potentials    over convex ones.}
\label{fig:inf_ns_r_latest}
\end{center}
\end{figure}

 Given the latest CMB constraints on $\ns$ and  especially the upper bound on the tensor-to-scalar ratio $r$ 
in (\ref{eq:CMB_r_obs}),  a number of 
famous models of inflation are strongly disfavoured. These include  the simplest quadratic chaotic potential $\f{1}{2} m^2\phi^2$, along with other monotonically increasing convex potentials, see figure \ref{fig:inf_ns_r_latest}. It is important to stress that  the entire family of  power law potentials $V(\phi) \propto \phi^p$ are now observationally ruled out as shown by the lime colour stripe in figure \ref{fig:inf_ns_r_latest}.

\bigskip

 Indeed,  CMB observations now  favour potentials exhibiting asymptotically flat plateau-like wings. In the literature of  inflationary model building \cite{Martin:2013tda}, there exist a number of plateau potentials that  satisfy the CMB constraints. 
These include single parameter potentials such as the  Starobinsky model  \cite{Starobinsky:1980te,Whitt:1984pd,Mishra:2018dtg}, and the non-minimally coupled Standard Model (SM) Higgs inflation  \cite{Fakir:1990eg,Bezrukov:2007ep,Mishra:2018dtg}, along with a number of important plateau  potentials with two parameters, such as  the T-model and E-model  $\alpha$-attractors  \cite{Kallosh:2013hoa,Kallosh:2013yoa}, and the D-brane  KKLT potential  \cite{Kachru:2003aw,Kachru:2003sx,Kallosh:2019hzo,Kallosh:2021mnu}  (where KKLT stands for Kachru-Kallosh-Linde-Trivedi). 

\subsection{Important plateau potentials as future CMB targets}
\label{sec:plateau_can}

\begin{figure}[htb]
\centering
\hspace{-0.4in}
\subfigure[]{
\includegraphics[width=0.51\textwidth]{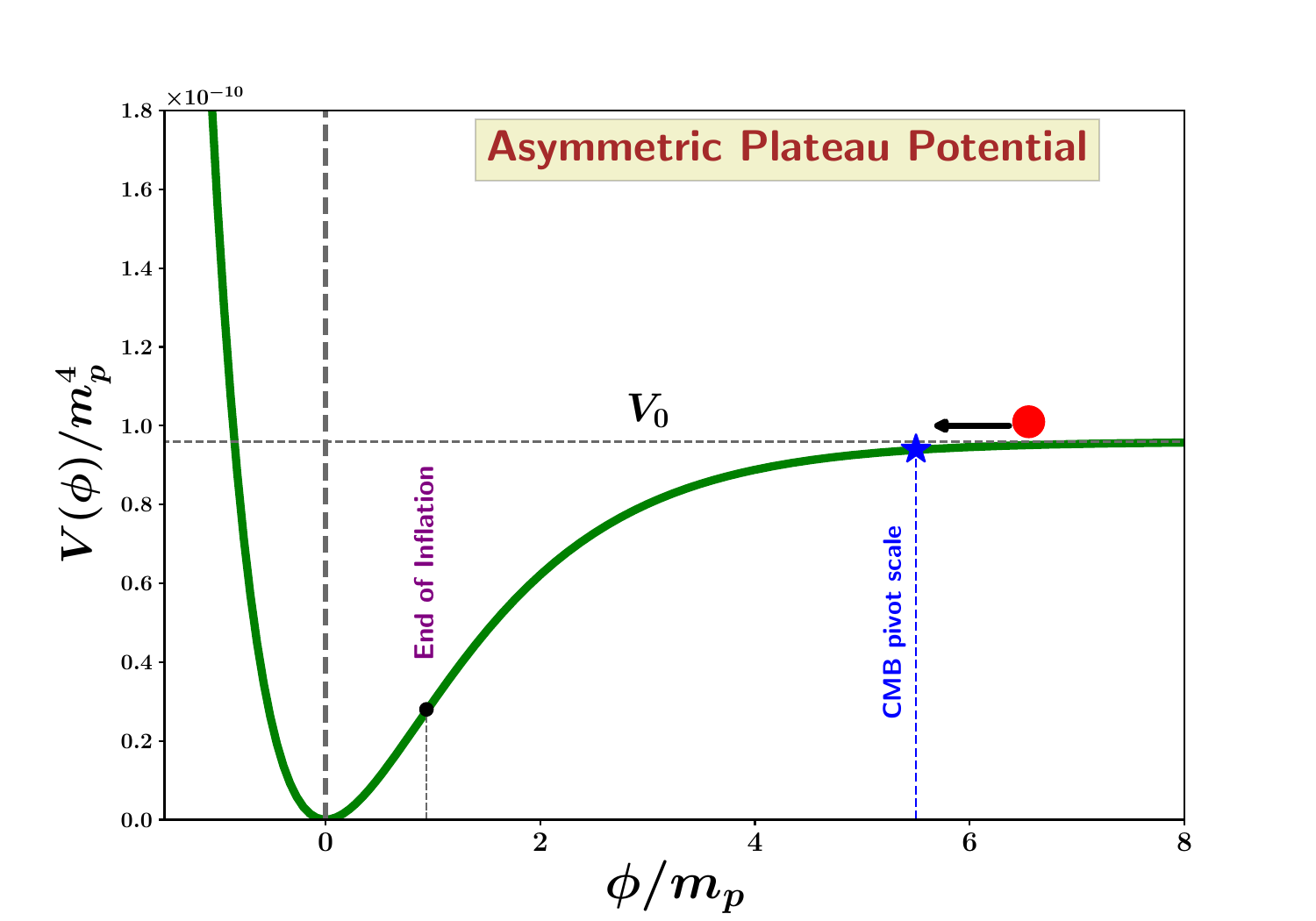}}\hspace{-0.4in}
\subfigure[]{
\includegraphics[width=0.51\textwidth]{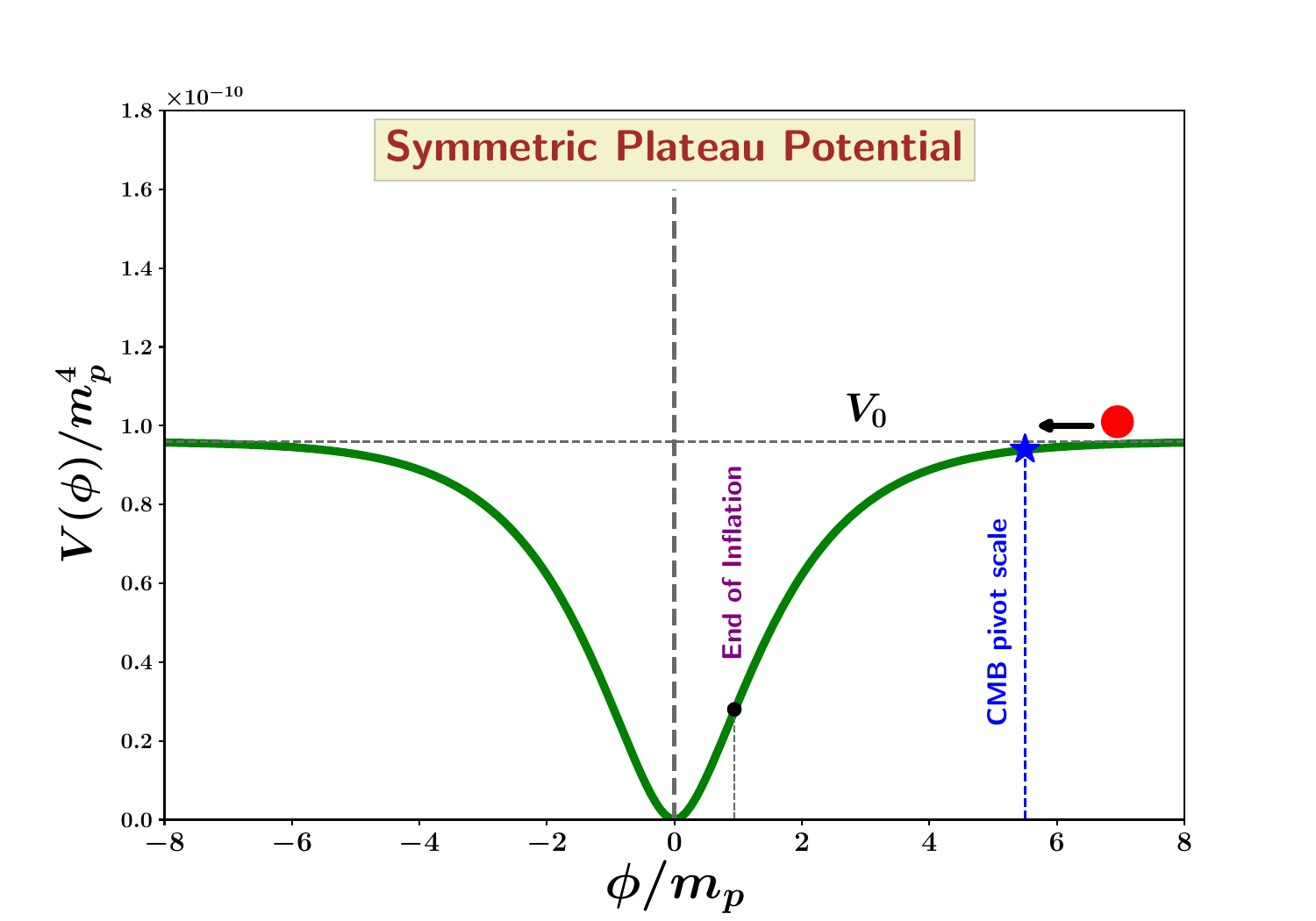}}
\caption{This figure schematically depicts asymptotically flat inflationary plateau potentials. The {\bf left panel} shows an asymmetric  potential featuring only a single plateau-like wing which supports slow-roll inflation. While the {\bf right panel} shows a symmetric  potential possessing two plateau-like wings, both of which can support slow-roll inflation.  Note that  for large field values, some  potentials  approach plateau behaviour exponentially, while others approach algebraically.} 
\label{fig:pot_plateau_toy}
\end{figure}

 An asymptotically flat potential has the general functional  form
 
 \beq
 V(\phi) = V_0 \, f\l(\lambda \f{\phi}{m_p}\r)~,
 \label{eq:pot_plateau_toy}
 \eeq 
 such that $V(\phi) \longrightarrow V_0$ at large field values $\lambda\phi \gg m_p$,
where $\lambda$ is a free parameter.  Such a potential, schematically represented in figure \ref{fig:pot_plateau_toy}, usually features one or two plateau-like wings  for large field values away from the minimum of the potential, depending upon whether the potential is asymmetric or symmetric.  Additionally, an asymptotically flat  potential might approach the plateau either exponentially or algebraically.  Below we briefly  discuss  a number of important plateau potentials in light of the latest CMB observations, keeping  one example of plateau models belonging to either of the following categories\footnote{  Our primary intention here is to provide a few examples of plateau models whose predictions can be accommodated by the latest data, rather than reviewing all available  plateau potentials in the inflationary literature. Similarly, we do not carry out a full parameter estimation program  and    likelihood comparison of different plateau models (which requires  MCMC analysis and is out of scope of this paper). We again stress that we have chosen one potential from each of the above categories just to provide an example for the reader. We do not claim that these potentials are more favourable than other models that have not been discussed in this section.}   - 

\begin{enumerate}
\item Symmetric or asymmetric plateau potentials.

\item Single or double parameter plateau potentials.

\item Potentials with exponential or algebraic approach to plateau. 

\end{enumerate}

\begin{enumerate}
\item \textbf{Starobinsky model}

The potential for Starobinsky inflation \cite{Starobinsky:1980te} in the Einstein frame, in
which the scalaron\footnote{ Note that Starobinsky inflation was originally formulated as a modified gravity theory  with  the Jordan frame Lagrangian $f(R) = R + R^2/6m^2$ which contains an additional scalar degree of freedom compared to general relativity. Upon conformal transformation of the Jordan frame metric, one can arrive at the Einstein frame Lagrangian where the extra scalar degree of freedom takes the form of  a canonical scalar field, known as the `scalaron'.} takes the form of a canonical scalar field minimally coupled to gravity,  is given by  \cite{Whitt:1984pd,Mishra:2018dtg}

\beq
V(\phi) =  V_0 \, \l( 1 - e^{-\f{2}{\sqrt{6}}\f{\phi}{m_p}} \r)^2~.
\label{eq:pot_star}
\eeq

This potential features a single parameter $V_0$, which is related to the scalaron mass  $m$ by 
$V_0 = \f{3}{4}m^2 m_p^2$ \cite{Mishra:2018dtg}, and whose value is completely fixed by the CMB normalization (\ref{eq:CMB_As_obs}). The prediction of $\lbrace \ns,r \rbrace$ for the Starobinsky potential  is given by 

\ber
\ns \simeq 1 - \f{2}{N_*} \, , ~~ r \simeq  \f{12}{N_*^2} \, , ~{\rm for}~ N_* \gg 1 \, ,
\label{eq:ns_r_star}
\eer
which is shown by black colour dots in figure \ref{fig:inf_ns_r_latest}. As per standard convention, the smaller and larger  black dots  represent    $\lbrace \ns,r \rbrace$ predictions of Starobinsky potential corresponding to $N_* = 50, \, 60$ respectively. It is important to note that  the predictions of Starobinsky inflation lie at the centre of the observationally  allowed region of $\lbrace \ns,r \rbrace$ (shown in grey in fig. \ref{fig:inf_ns_r_latest}), making it  one of the most popular inflationary models at present.

\item \textbf{Non-minimally coupled Standard Model (SM) Higgs inflation}
  
  The SM Higgs inflation was originally formulated \cite{Fakir:1990eg,Bezrukov:2007ep} in the Jordan frame 
where the Higgs field is non-minimally coupled to the scalar  (Ricci) curvature of space-time.
One can however transform it to the Einstein frame by a conformal transformation of the 
metric \cite{Mishra:2018dtg,Bezrukov:2007ep} so that  the Einstein frame action  describes   a canonical scalar 
field minimally coupled to gravity. The corresponding SM Higgs inflaton potential, for large field values, takes the form 
  
\beq
V(\phi) \simeq  V_0 \, \l( 1 - e^{-\f{2}{\sqrt{6}}\f{\vert \phi\vert}{m_p}} \r)^2~.
\label{eq:pot_SM_nm_Higgs}
\eeq

The above potential also  features a single parameter $V_0$, which is related to the strength of the non-minimal coupling $\xi$  by $V_0=\frac{\lambda m_p^4}{4\xi^2}$ as shown in  \cite{Bezrukov:2007ep,Mishra:2018dtg}, where $\lambda = 0.1$ is the SM Higgs self-coupling. Its  value is completely fixed by the CMB normalization (\ref{eq:CMB_As_obs}).  Given the similarity in the functional form of  (\ref{eq:pot_star}) and  (\ref{eq:pot_SM_nm_Higgs}), the 
prediction  of  the non-minimally coupled SM Higgs inflation for $\lbrace \ns,r \rbrace$   is similar to that of 
Starobinsky inflation\footnote{Note that while   the functional  form of  the Einstein frame potentials for Starobinsky inflation  (\ref{eq:pot_star}) looks similar to that of the non-minimally coupled SM Higgs inflation (\ref{eq:pot_SM_nm_Higgs}), however, the Starobinsky  potential is asymmetric, while the Higgs potential is symmetric \cite{Mishra:2018dtg}. Similarly, the predictions of $\lbrace \ns,r \rbrace$ for both the models can in principle be different, depending upon the details of reheating mechanism \cite{Bezrukov:2011gp}.}, given by  (\ref{eq:ns_r_star}).

\item \textbf{T-model $\alpha$-attractor}

The T-model $\alpha$-attractor potential \cite{Kallosh:2021mnu}  is a symmetric plateau potential of  the functional form 

\beq
V(\phi)  =V_0 \, \tanh^{p}{\l(\lambda \f{\phi}{m_p}\r)} \, , 
\label{eq:pot_Tmodel}
\eeq
 where $p$ is a positive real number. For a given value of $p$, the T-model potential has two parameters, namely $V_0$ and $\lambda$. As usual, the value of $V_0$ is fixed by the CMB normalization (\ref{eq:CMB_As_obs}) while   $\lambda$ determines\footnote{In the T-model potential,  $\lambda$ is related to the $\alpha$ parameter of $\alpha$-attractors \cite{Kallosh:2013yoa} by $\lambda^2 = \f{1}{6\alpha}$.} the predicted value of  $\lbrace \ns,r \rbrace$ for this potential.

\begin{figure}[htb]
\begin{center}
\includegraphics[width=0.8\textwidth]{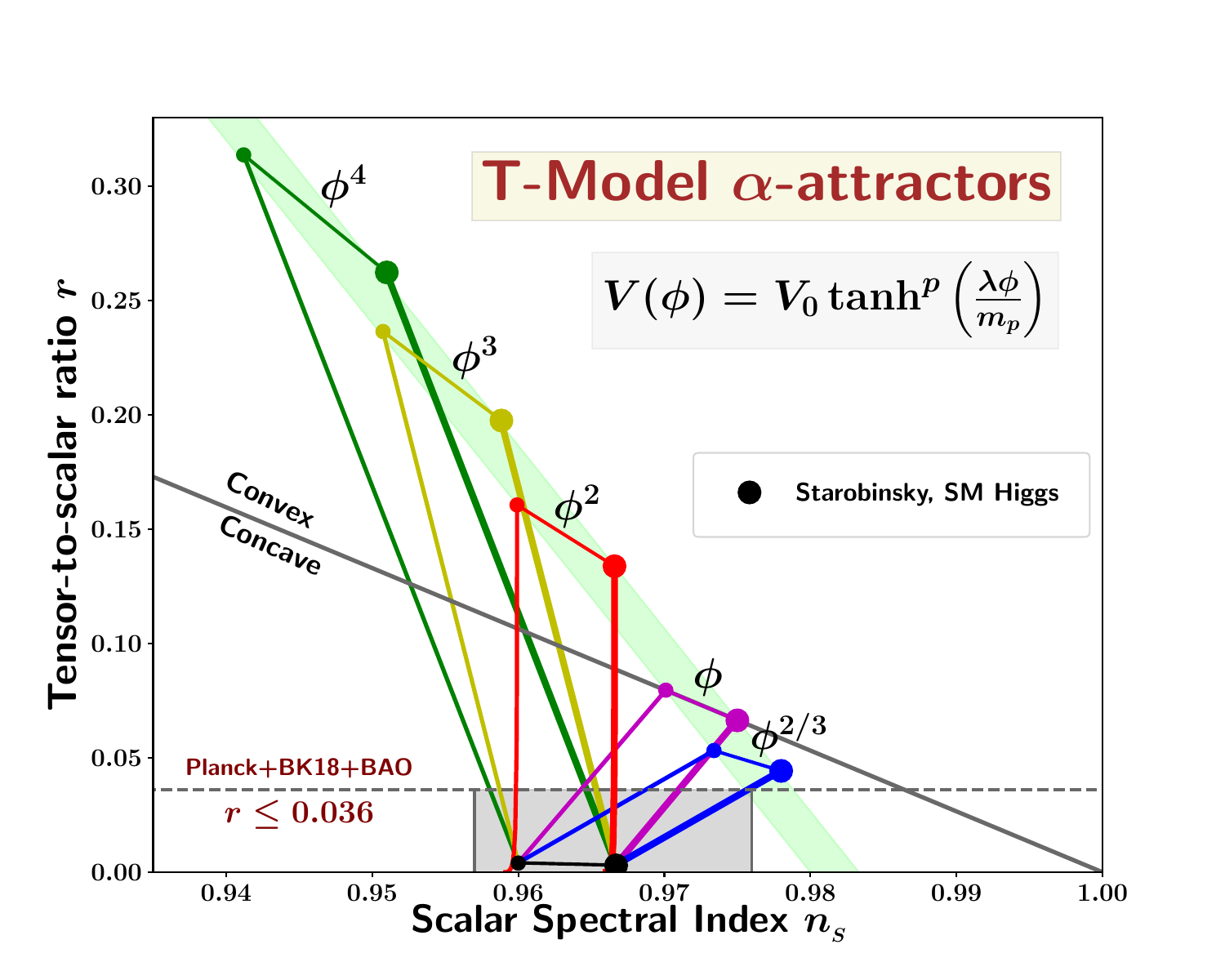}
\caption{This figure is a plot of  the  tensor-to-scalar ratio $r$, versus  the  scalar spectral index $\ns$, for the  T-model  $\alpha$-attractor (\ref{eq:pot_Tmodel})  (the thinner and thicker   curves correspond to $N_* = 50,\, 60$ respectively). The latest CMB 2$\sigma$ bound 
$0.957 \leq \ns \leq 0.976$
 and
 the upper bound on the tensor-to-scalar ratio  $r\leq 0.036$ are
 indicated by the shaded grey colour region. It is important to note that upon increasing the value of $\lambda$, the predictions of T-model potential (\ref{eq:pot_Tmodel}) for different values of $p$  converge towards the cosmological attractor at the centre, described by equation (\ref{eq:ns_r_Tmodel_att}).  For $\lambda \gtrsim 0.1$, the predictions of T-model become compatible with the latest CMB $2\sigma$ bound (see table \ref{table:app1})}.
\label{fig:inf_ns_r_Tmodel}
\end{center}
\end{figure}

\medskip
 Predictions of the  simplest T-model potential, which  corresponds to $p=2$ in equation (\ref{eq:pot_Tmodel}), are shown by the red colour curves in figure \ref{fig:inf_ns_r_latest}. As we vary  value of $\lambda$ from $\lambda: 0\longrightarrow \infty$, the $r$ versus $\ns$ values trace out  continuous  curves   in figure \ref{fig:inf_ns_r_latest} (the thinner and thicker red curves correspond to $N_* = 50,\, 60$ respectively). We notice  that  in one limit, namely $\lambda \ll 1$, the CMB predictions for   $\lbrace \ns,r \rbrace$ of the T-model matches with that of the quadratic $\phi^2$ potential    owing to the fact that the potential (\ref{eq:pot_Tmodel}) for $p=2$ behaves like $V(\phi) \propto \phi^2$ for $\lambda \phi \ll m_p$. In fact this is true in general  for the T-model potential with any value of $p$ in  (\ref{eq:pot_Tmodel}),  leading to the behaviour $V(\phi) \propto \phi^p$  for $\lambda\phi \ll m_p$, which is demonstrated in figure \ref{fig:inf_ns_r_Tmodel}. However in the opposite limit, namely $\lambda \geq 1$ (which results in $\exp{\l(\lambda\phi_*/m_p\r)}  \gg 1$), the predictions of T-model become  \cite{Kallosh:2021mnu} 
  
\beq
\ns \simeq 1 - \f{2}{N_*} \, , ~~ r \simeq  \f{2}{\lambda^2 N_*^2} \, , ~{\rm for}~ N_* \gg 1\, ,
\label{eq:ns_r_Tmodel_att}
\eeq  

which is independent of $p$ (see \cite{Mishra:2021wkm}). 
Due to this property these models are called `cosmological attractors' in \cite{Kallosh:2021mnu}.  In section \ref{sec:non_can_powerlaw}, we will demonstrate that the $r$ versus $\ns$  flow lines of  monomial potentials $V(\phi) \sim \phi^p$ in the non-canonical framework  display striking similarities (see figure \ref{fig:inf_ns_r_nc_powerlaw}) with the predictions of T-model shown in figure \ref{fig:inf_ns_r_Tmodel}.

\item \textbf{E-model $\alpha$-attractor}

The E-model $\alpha$-attractor potential \cite{Kallosh:2021mnu}  is an asymmetric plateau potential of  the functional form 

\beq
V(\phi) =  V_0 \, \l( 1 - e^{-\lambda\f{\phi}{m_p}} \r)^{p} \, , 
\label{eq:pot_Emodel}
\eeq
 where $p$ is a positive real number. For $p=2$ and $\lambda = 2/\sqrt{6}$, the E-model potential coincides with Starobinsky potential\footnote{In the E-model potential,  $\lambda$ is related to the $\alpha$ parameter of $\alpha$-attractors \cite{Kallosh:2013yoa} by $\lambda^2 = \f{2}{3\alpha}$}. In the limit $\lambda \geq 1$ (which results in $\exp{\l(\lambda\phi_*/m_p\r)}  \gg 1$), the predictions of E-model become  \cite{Kallosh:2021mnu} 

\beq
\ns \simeq 1 - \f{2}{N_*} \, , ~~ r \simeq  \f{8}{\lambda^2 N_*^2} \, , ~{\rm for}~ N_* \gg 1 \, ,
\label{eq:ns_r_Emodel_att}
\eeq  

which  again is  independent of $p$ (see \cite{Mishra:2021wkm}). The CMB predictions $\lbrace \ns,r \rbrace$   of  E-model potential   are shown by the cyan colour curves in figure \ref{fig:inf_ns_r_latest} for the case $p=2$. (the thinner and thicker cyan curves correspond to $N_* = 50,\, 60$ respectively).

\bigskip 

In all of the above models, the potential approaches plateau behaviour exponentially. In the following we briefly discuss KKLT potential which approaches plateau behaviour algebraically.

\item \textbf{D-brane KKLT potential}

The D-brane KKLT inflation~\cite{Kachru:2003aw,Kachru:2003sx} potential (which has the same functional form as the polynomial $\alpha$-attractor model~\cite{Kallosh:2022feu}) is given by~\cite{Kallosh:2019hzo,Kallosh:2021mnu}
\beq
V(\phi) = V_0 \, \f{\phi^n}{\phi^n+M^n}\, , 
\label{eq:pot_KKLT}
\eeq
where  $n$ is a positive integer and  $M$ is a fundamental  scale of the theory. This is a symmetric plateau potential, like the T-model $\alpha$-attractor (\ref{eq:pot_Tmodel}).  However the KKLT potential approaches  plateau behaviour (saturates) algebraically, in contrast to  the exponential approach to plateau behaviour exhibited  by the T-model potential. 

\begin{figure}[htb]
\begin{center}
\includegraphics[width=0.8\textwidth]{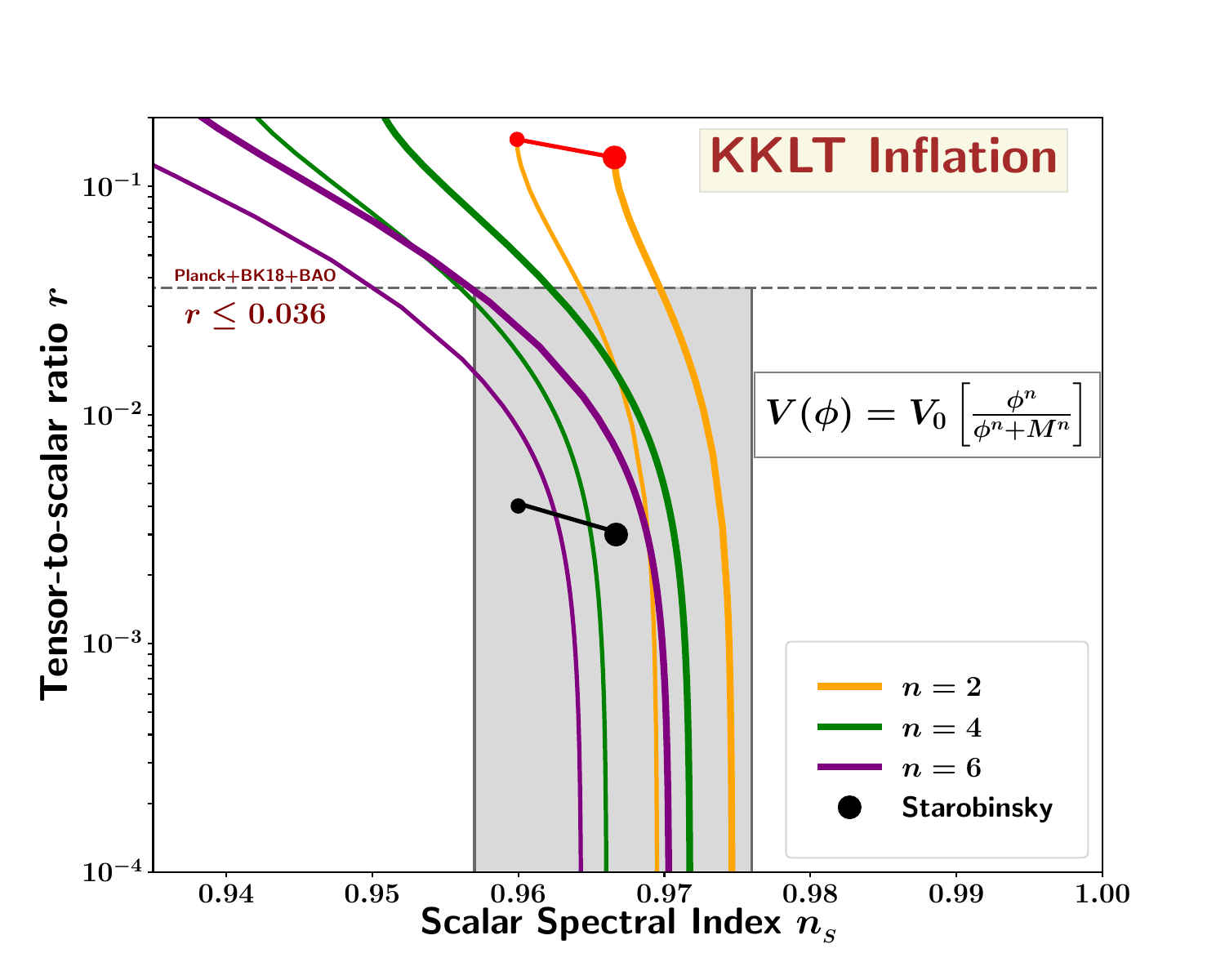}
\caption{This figure is a plot of  the  tensor-to-scalar ratio $r$ versus  the  scalar spectral index $\ns$    for  KKLT inflation (\ref{eq:pot_KKLT})  (the thinner and thicker   curves correspond to $N_* = 50,\, 60$ respectively).    The latest CMB 2$\sigma$ bound 
$0.957 \leq \ns \leq 0.976$
 and
 the upper bound on the tensor-to-scalar ratio  $r\leq 0.036$ are
 indicated by the shaded grey colour region. Upon decreasing the value of parameter $M$ in the KKLT potential (\ref{eq:pot_KKLT}), the tensor-to-scalar ratio $r$ decreases and hence the model satisfies the CMB bound for smaller values of $M$ (see table \ref{table:app3}).}
\label{fig:inf_ns_r_KKLT}
\end{center}
\end{figure}

In the limit $M\gg m_p$,   the potential asymptotes to a monomial  power law form $V(\phi) \propto \phi^n$, the predictions of which are strongly disfavoured by observations. In the opposite limit, namely $M\ll m_p$, the potential is plateau-like, $V(\phi) \longrightarrow V_0$, whose predictions satisfy the CMB data very well. Figure \ref{fig:inf_ns_r_KKLT} shows the $r$ versus $\ns$ plot for KKLT potential (\ref{eq:pot_KKLT}) for three different values of $n$ (the thinner and thicker  curves correspond to $N_* = 50,\, 60$ respectively).  From this figure, it is easy to infer that  the  predictions of KKLT potential (\ref{eq:pot_KKLT}) for $\lbrace \ns, r \rbrace$  do not  reach a common attractor regime for different values of $n$ (in contrast to those of the T-model and E-model), rather they  cover a large horizontal  portion of the observationally allowed region of $\ns$, as highlighted in \cite{Kallosh:2019hzo,Kallosh:2021mnu}. 

\end{enumerate}

\bigskip

Note that the values of  the parameter $\lambda$ in the T-model and E-model potentials, and the parameter $M$ in the KKLT potential, for which these models become compatible with the latest CMB data are discussed in appendix \ref{sec:app_bounds}.

\medskip

  In this section, we discussed the predictions for  $\lbrace \ns,r \rbrace$  of a number of plateau potentials  
which are consistent with current CMB data and are
interesting  target candidates for next generation CMB missions\footnote{In passing, we would  again like to stress that the set of plateau potentials discussed in section \ref{sec:plateau_can} is far from being exhaustive of all the inflationary models that are consistent with the CMB data, for which we refer the reader to more extensive  literature on inflationary 
models \cite{Martin:2013tda}.}. 
Despite their excellent agreement with observations, plateau potentials may face some
theoretical shortcomings which relate to the issue of
initial conditions. Although a large range of initial  field values and field velocities 
$\lbrace \phi_i, \dot{\phi_i}\rbrace$ results in adequate inflation for a plateau potential 
 \cite{Mishra:2018dtg}, the small height of the plateau $V_0 \leq 10^{-10}\, m_p^4$ does not allow for the equipartition of initial  kinetic, potential and gradient terms in the Lagrangian density \cite{Ijjas:2013vea}. 
Similarly the presence of a non-negligible  initial positive spatial curvature term
  might also prove problematic for plateau inflation especially if $|\rho_K^{\rm initial}| \gg V_0 \sim 10^{-10}\, m_p^4$ (see  \cite{Ijjas:2013vea}).  Additionally, plateau potentials predicting a small value of the tensor-to-scalar ratio might exhibit relatively large running of the scalar spectral index $\ns$ (see \cite{Easther:2021rdg}).

These   potentially important issues associated with plateau potentials  
can be addressed in two distinct ways\footnote{Note that Starobinsky and SM Higgs inflation were originally formulated in their respective  Jordan frames. However it is always possible to go to the Einstein frame by a conformal transformation of the 
metric, in which the action takes the form of a single canonical scalar field minimally coupled to gravity. Additionally, the T-model and E-model $\alpha$-attractors, as well as the D-brane KKLT inflation,  can  be formulated  in a framework in
which the Lagrangian density  consists of   a non-canonical kinetic term featuring a pole and a monomial potential  \cite{Kallosh:2021mnu}. However, it is possible to go to the  canonical framework by a mere field redefinition,  such that the kinetic term is canonical and the potential is  asymptotically flat, as given by  (\ref{eq:pot_Tmodel}), (\ref{eq:pot_Emodel}), (\ref{eq:pot_KKLT}).
}: 

(i) Modifying the form of the potential $V(\phi)$ by making it convex at large
values of $\phi$ while preserving its plateau-like form at moderate values of $\phi$
and keeping the canonical nature of the Lagrangian intact.

(ii) By exploring convex potentials, including $V \propto \phi^p$, in a non-canonical framework
in which the Lagrangian density has the form ${\cal L} = X^\alpha - V(\phi)$.

The first of these possibilities will be examined in the next section while non-canonical scalars will be discussed in
section \ref{sec:inf_noncan}.

 \section{Inflation with the Margarita potential}
\label{sec:margarita}  

Most of the models discussed in the previous sections had plateau-like wings which prevented inflation from
occuring at Planck scale values of the inflationary potential, and hence disfavoured the presence of 
equipartition initial conditions with ${\dot\phi}^2 \sim V(\phi) \sim \mpl^4$.
This minor flaw of plateau potentials is alleviated in a new family of potentials which we call the
Margarita family.
The Margarita potential introduces an intermediate flat wing on a potential which thereafter increases
monotonically with $\phi$. It therefore has properties similar to both plateau and monomial potentials.
The Margarita potential allows for the presence of equipartition initial conditions while at the same
time providing excellent agreement with current CMB constraints.

Margarita-type potentials\footnote{The Margarita potential is inspired by the shape of the Margarita cocktail glass.
It was originally introduced in \cite{Bag:2017vjp} to facilitate a period of  transient acceleration which
gave rise to dark energy. In this
 context, the Margarita potential featured a monotonically growing steep exponential  wing at large field values. 
This allowed the small current value of the DE density to commence
 from a large basin of initial conditions.} have the following general features --
\begin{enumerate}
\item
Monotonically growing wings support inflation  at large values of the  inflaton field.
\item An oscillatory quadratic asymptote, $V \propto \phi^2$, exists
at small values of the inflaton field.
\item An intermediate plateau-like wing joins 1 and 2. 
\end{enumerate}

We propose the following simple functional form for Margarita-type potentials

\beq
V(\phi)=V_b\l(\lambda_1,\phi\r) V_c\l(\lambda_2,\phi\r),~~ \lambda_1 \gg \lambda_2~.
\label{eq:pot_margarita_gen}
\eeq
Here $V_b(\phi)$ is an asymptotically flat plateau-like potential with height $V_0$, while $V_c(\phi)$ is the correction to the plateau-like wing at large inflaton field values. 
The two free parameters $\lambda_1$ and $\lambda_2$ in (\ref{eq:pot_margarita_gen}) satisfy
 $\lambda_1 \gg \lambda_2$. $V_c(\phi)$ increases monotonically for $\phi/m_p\gg \f{1}{\lambda_2}$ while $V_c(\phi=0)=1$. 
In this work we shall choose $V_c(\phi)$ such that $V_0 V_c(\phi)\sim V_0+\f{1}{2}m_2^2\phi^2$ for $\phi/m_p\ll \f{1}{\lambda_2}$,  where $m_2$ depends on $\lambda_2$.

\begin{figure}[htb]
\begin{center}
\includegraphics[width=0.8\textwidth]{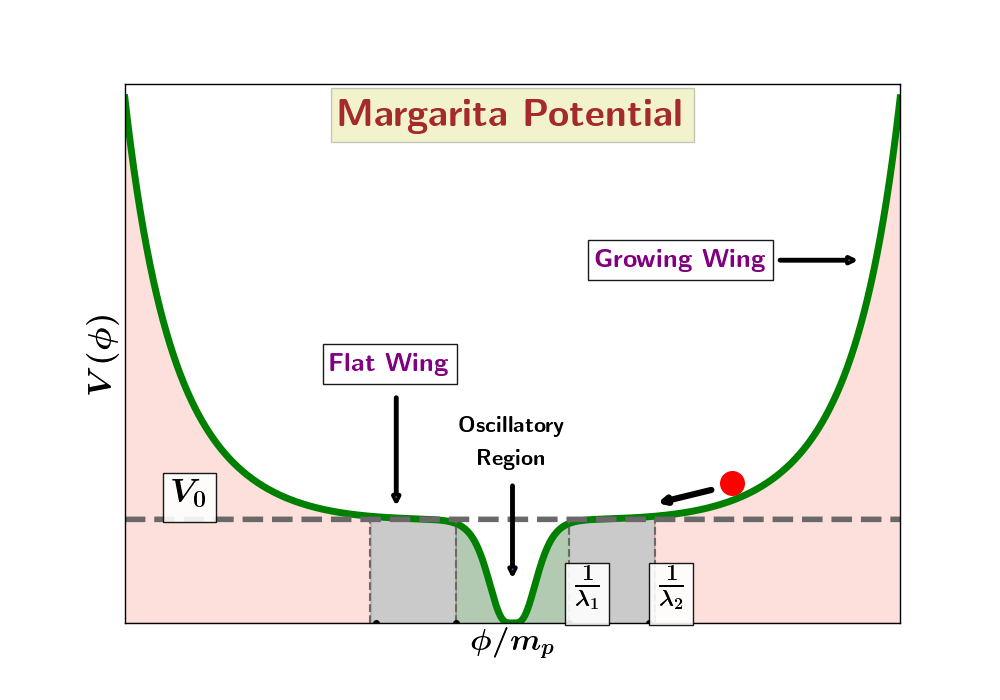}
\caption{This figure schematically illustrates a typical Margarita potential (\ref{eq:pot_margarita_T_exp}) with $\lambda_1 \gg \lambda_2$. For $\lambda_2|\phi| \gg m_p$, the potential exhibits  monotonically growing convex inflationary  wings (red shade). For small field values, namely $\lambda_1|\phi| \ll m_p$, the potential has a minimum (green shade). While for intermediate field values  $\f{1}{\lambda_1} < \f{\phi}{m_p} < \f{1}{\lambda_2}$, the 
potential displays plateau-like behaviour (grey shade).}
\label{fig:pot_margarita_gen}
\end{center}
\end{figure}

As a result, a generic Margarita  potential will
exhibit the following three asymptotic branches (see figure \ref{fig:pot_margarita_gen}) --

\ber
\mbox{Growing wing:} \quad V(\phi) &\simeq& {\rm Monotonic}~,  \quad \frac{|\phi|}{m_p} \gg \frac{1}{\lambda_2}\, ,
\label{eq:pot_mar1}\\
\mbox{Flat wing:} \quad V(\phi) &\simeq& V_0+\frac{1}{2}m_2^2\phi^2\, ,\quad \frac{1}{\lambda_1}\ll \frac{|\phi|}{m_p} \ll \frac{1}{\lambda_2}\, ,
\label{eq:pot_mar2}\\
\mbox{Oscillatory region:} \quad V(\phi) &\simeq& \frac{1}{2}m_1^2\phi^2\, , \quad \frac{|\phi|}{m_p} \ll \frac{1}{\lambda_1}\, ,
 \label{eq:pot_mar3}
\eer 
where the mass parameters  $m_1$ and $m_2$ depend on $\lambda_1$ and $\lambda_2$ respectively. Figure \ref{fig:pot_margarita_gen} illustrates the three asymptotic branches of the Margarita potential.

We first demonstrate the behaviour of Margarita potential by assuming $V_b$ to be
 the T-model  based plateau potential \cite{Kallosh:2013hoa,Kallosh:2013yoa}. 

\subsection{T-model based exponential Margarita Potential}

The Margarita potential in this case is given by 
\beq
V(\phi)=V_0 \tanh^2{\l(\f{\lambda_1\phi}{m_p}\r)} \cosh{\l(\frac{\lambda_2\phi}{m_p}\r)}~.
\label{eq:pot_margarita_T_exp}
\eeq 
where
the base potential 
\beq
V_b(\phi)= V_0 \tanh^2{\l(\f{\lambda_1\phi}{m_p}\r)} ~,
\label{eq:pot_plateau_Tmodel}
\eeq
is the familiar   T-model $\alpha$-attractor  potential whose $r$ versus $n_{_S}$ plot is shown by the red colour curve  in figure \ref{fig:inf_ns_r_Tmodel}. 
The correction potential, $V_c(\phi)=\cosh{\l(\frac{\lambda_2\phi}{m_p}\r)}$, has exponential asymptotes for large field values $\lambda_2\phi \gg m_p$.

\begin{figure}[htb]
\begin{center}
\includegraphics[width=0.8\textwidth]{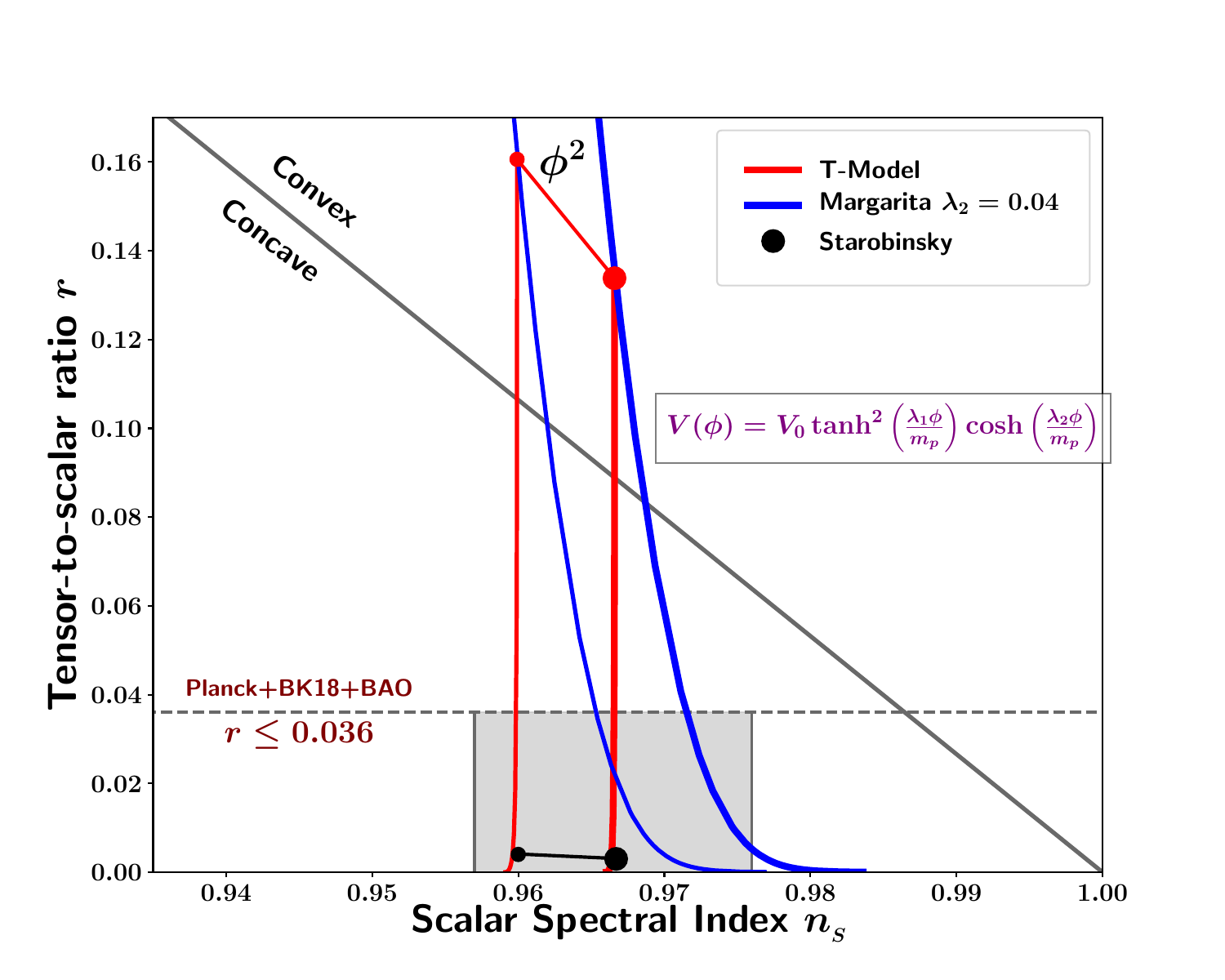}
\caption{This figure compares the $r$ versus $n_{_S}$ plot of the margarita potential (\ref{eq:pot_margarita_T_exp})   with its asymptotically flat T-model  base potential (\ref{eq:pot_plateau_Tmodel}) for  $\lambda_2 =0.04$ (the tensor-to-scalar ratio $r$ decreases upon increasing the value of $\lambda_1$  and becomes compatible with the CMB bound $r \leq 0.036$  for $\lambda_1 \gtrsim 0.1$}). As usual, the thinner and thicker curves correspond to $N_* =50, 60$ respectively. 
\label{fig:margarita_Tmodel_ns_r}
\end{center}
\end{figure}

In this  Margarita potential (\ref{eq:pot_margarita_T_exp}), the exponential  correction  arising from 
$V_c(\phi)=\cosh{\l(\frac{\lambda_2\phi}{m_p}\r)}$  to the base plateau potential modifies the nature of the $r$ versus $n_{_S}$ 
plot, shifting this curve towards higher values of $n_{_S}$ for identical values of $r$.  Hence the $\lbrace \ns,r \rbrace$ 
predictions of the Margarita potential are, in principle, distinguishable from those of the base T-model potential. 
This shift is dependent upon the value of the correction parameter $\lambda_2$.  For $\lambda_2 = 0$, 
one arrives at the base T-model potential while larger values of $\lambda_2$, give rise to a larger shift in
the value of $\ns$ with respect to the base potential.   Note that we assume $\lambda_2 < \sqrt{2}$ in order to realise inflation in the exponential asymptote to $V(\phi)$ at early times.   We find that the Margarita potential  (\ref{eq:pot_margarita_T_exp}) satisfies the CMB constraints on $\ns$ for $\lambda_2 \leq 0.1$ (while the CMB upper bound on $r$ can be satisfied for suitable range of values of $\lambda_1$, as in the case of T-model). Figure \ref{fig:margarita_Tmodel_ns_r} shows the $r$ versus $n_{_S}$ plot of this Margarita potential  for $\lambda_2 = 0.04$. 

\subsection{Implications for initial conditions}

  As discussed in section \ref{sec:plateau_can},  although recent  CMB observations appear to
 favour asymptotically flat plateau potentials, the latter could face some theoretical shortcomings relating  to the issue of initial conditions. In particular, owing to the fact that the height of the plateau saturates to a small value $V_0 \leq 10^{-10}~m_p^4$ for large field values $\phi \gg m_p$, an equipartition of the initial inflaton  kinetic energy density  $\rho_{\rm KE} = \f{1}{2}\dot{\phi}^2$, potential energy density  $\rho_{\rm V} \simeq V_0$, and gradient  term $\rho_{\rm grd} \simeq \f{1}{2}(\nabla\phi)^2$ will  not be possible. Moreover, the possible presence of a large initial curvature
 density $\rho_K = -3\,m_p^2/a_i^2$ (corresponding to positive spatial curvature, $K=+1$)  might also 
be problematic for inflation.

\begin{figure}[htb]
\centering
\hspace{-0.4in}
\subfigure[]{
\includegraphics[width=0.48\textwidth]{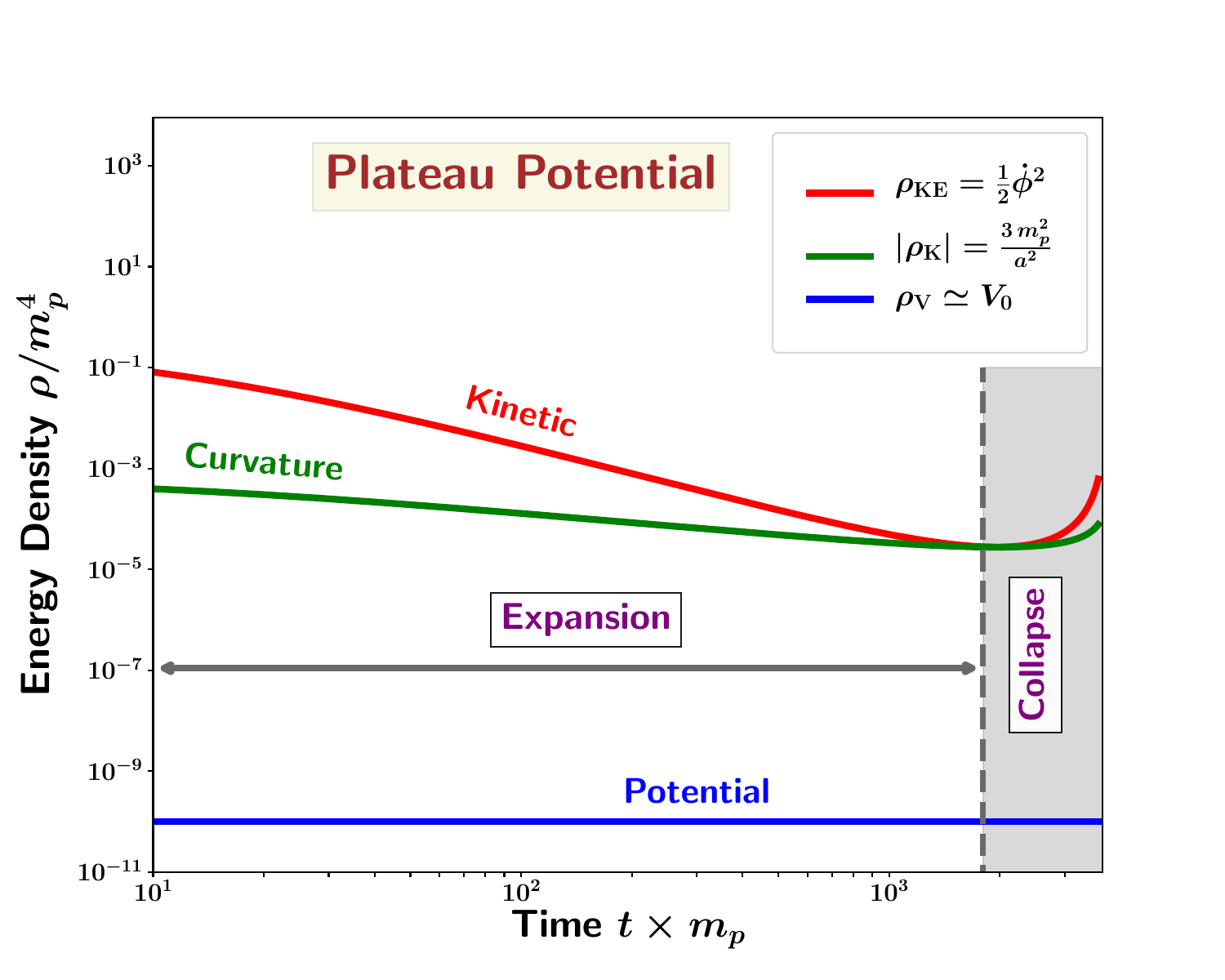}}\hspace{-0.4in}
\subfigure[]{
\includegraphics[width=0.48\textwidth]{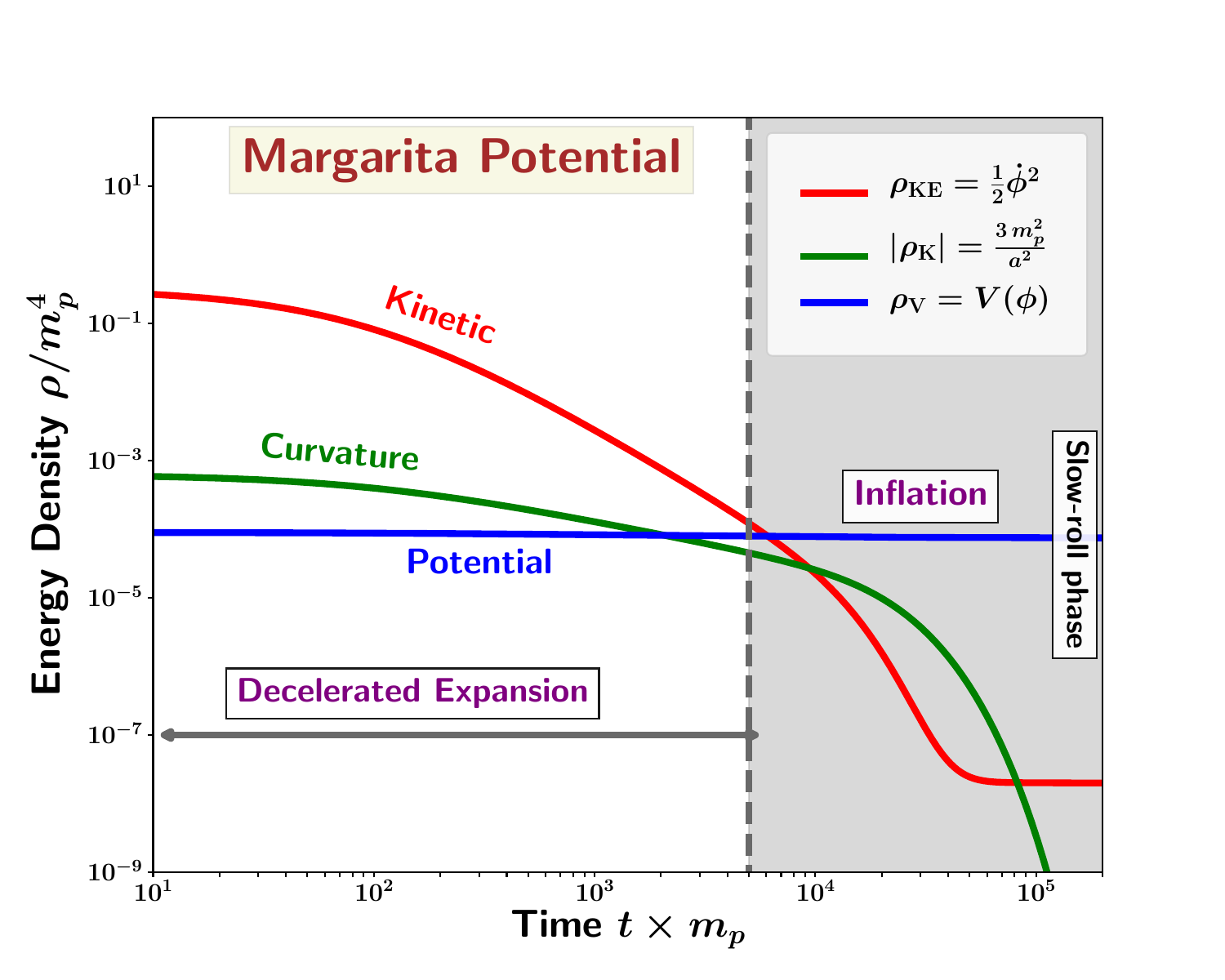}}
\caption{This figure illustrates the implications of a relatively large initial
curvature density $\rho_K = -3\,m_p^2/a_i^2$ (corresponding to a positive spatial curvature, $K=+1$) 
at the commencement of  inflation. The {\bf left panel}  shows the density  evolution of the inflaton kinetic term $\rho_{\rm KE}$ (red), potential term $\rho_V$ (blue) and the modulus of the spatial curvature term $|\rho_K|$ (green) 
for a plateau potential with $V(\phi) \simeq V_0 = 10^{-10}\,m_p^4$. The {\bf right panel} shows the same for
 the Margarita potential (\ref{eq:pot_margarita_T_exp}). From the left panel, we notice that for a plateau potential, even if
$\rho_{\rm KE} \gg |\rho_K|$ initially, an expanding universe collapses 
even before inflating due to the small value of $\rho_V$ in plateau potentials. 
By contrast the right panel shows that even though the initial value of $\rho_V$ is smaller than the curvature term, the universe 
keeps expanding and eventually begins to inflate. If one commences from an initial equipartition configuration 
in which all the three densities are comparable, then realising inflation in a Margarita potential  is extremely easy.} 
\label{fig:rho_plateau}
\end{figure}

               To emphasise this point, let us imagine that we set initial conditions for inflation closer to the Planck scale, \textit{i.e} $\rho_{\phi_i} \simeq  \, 0.1\,m_p^4$. Suppose the height of the plateau is $V_0 \simeq 10^{-10}~m_p^4$. In this case, starting from an initial state where the kinetic and curvature densities are of similar magnitude, namely $\rho_{\rm KE} \simeq \rho_K \simeq \rho_{\phi_i}$, we notice that the universe begins to collapse before inflation begins. This has been demonstrated in the left panel of  figure \ref{fig:rho_plateau} which shows that even an initial  curvature density as small as $1\%$ of the initial  inflaton energy density leads to the eventual collapse of an initially expanding universe. In fact the initial curvature density must be extremely sub-dominant compared to the initial inflaton energy density in order to prevent the collapse. A preliminary numerical analysis in a forthcoming paper \cite{s_v_t_curvature} shows that  $\rho_K <  10^{-6}~\rho_{\phi_i}$  in order for inflation to begin. Hence, unless the spatial curvature is negative\footnote{Note that since inflation is supposed to address the flatness problem, we do not assume the spatial curvature of the universe to be zero initially.}  or,  it is positive with a
 negligibly small  density, plateau potentials are prone to the fine tuning  of initial conditions.

However both the equipartition problem and the positive spatial curvature issue can be successfully addressed by a
 Margarita potential with monotonically growing inflationary wings. This has been demonstrated in the right panel of figure \ref{fig:rho_plateau}. We notice that, starting close to the Planck scale\footnote{We keep the initial inflaton density $\rho_{\phi_i}$ to be one order smaller than the Planck scale in order to justify the usage of Einstein's GR in our numerical simulations.} with  a relatively large  initial curvature density, the universe proceeds to inflate successfully and the collapse scenario can be completely avoided.   It is easy to see that the equipartition condition can be easily satisfied in this case. Hence Margarita 
potentials could play a crucial role in addressing the initial (positive) curvature problem, as well as facilitating 
the equipartition of initial energy densities of different components, which were amongst the primary goals of inflation in  its original formulation \cite{Guth:1980zm, Linde:1981mu}.

\subsection{General recipe for constructing a Margarita potential}
\index{margarita}

The Margarita `cocktail glass'
 shape of the potential in figure \ref{fig:pot_margarita_gen} is quite general and not restricted to 
the potential (\ref{eq:pot_margarita_T_exp}) discussed previously.

Indeed a Margarita potential $V(\phi)$ can easily be constructed using the prescription discussed earlier, namely:

\begin{itemize}

\item Multiply a plateau-like base potential $V_b(\phi)$ with a convex inflationary potential $V_c(\phi)$ to make
the Margarita potential $V(\phi)$:

\beq
V(\phi) = V_b(\phi) \times V_c(\phi)
\eeq

\item Here $V_b(\phi)$ must have the following properties:

\begin{enumerate}

\item It vanishes at the origin $V_b(\phi=0) = 0$.

\item Its asymptotically flat: $V_b(\phi \to \pm\infty) = V_0$.

\end{enumerate}

Examples of some widely studied plateau potentials are:

1. The KKLT potential \cite{Kallosh:2019hzo,Kallosh:2021mnu}
\beq
V_b(\phi)=V_0\frac{\phi^n}{\phi^n+M^n}~.
\label{eq:base_KKLT}
\eeq

2. The T-model $\alpha$-attractor \cite{Kallosh:2019hzo,Kallosh:2021mnu}
\beq
V_b(\phi)=V_0\tanh^{2n}\left(\frac{\alpha\phi}{m_p}\right)~.
\label{eq:base_alpha}
\eeq

3. The (non-minimally coupled) standard model Higgs potential in the Einstein frame
\beq
V(\phi) \simeq  V_0 \, \l( 1 - e^{-\f{2}{\sqrt{6}}\f{\vert \phi\vert}{m_p}} \r)^2~.
\label{eq:pot_SM_nm_Higgs1}
\eeq

\item The convex  inflationary potential  function $V_c(\phi)$ should be dimensionless and preferably 
symmetric about the origin so that $V_c(\phi) = V_c(-\phi)$
and $V_c(\phi=0) = 1$.

\end{itemize}

\begin{table*}[tbh!]
\begin{center}
\begin{minipage}[h]{0.9\linewidth} \mbox{} \vskip 18pt
\begin{tabular}{llll}
\hline
\\
{\bf Plateau potential}  & {\bf Convex potential}  & {\bf Margarita potential} \\

\\
 {\large $V_b(\phi)$} &  {\large $~V_c(\phi)$} & {\large $~V(\phi)
=V_b(\phi)\times V_c(\phi)$}\\
\\\hline
& \\
$V_0\,\tanh^{2n}\left({\alpha\phi}\right)$ &  $\cosh{\left({\beta\phi}\right )}$ & 
$V_0\,\tanh^{2n}\left({\alpha\phi}\right)\cosh{\left({\beta\phi}\right )}$\\
& \\
$V_0 \,\frac{\phi^n}{\phi^n+M^n}$ &  $1+\alpha\phi^2$ & $V_0\,\frac{\phi^n}{\phi^n+M^n}\,(1+\alpha\phi^2)$
& \\
$V_0 \, \l( 1 - e^{-\alpha\f{\vert \phi\vert}{m_p}} \r)^2$ & $e^{\beta\phi^2}$ & $V_0 \, \l( 1 - e^{-\alpha\f{\vert \phi\vert}{m_p}} \r)^2e^{\beta\phi^2}$
& \\
& \\
\hline
\end{tabular}
\caption{This table lists three base potentials $V_b(\phi)$ which are asymptotically flat and three
 convex potentials $V_c(\phi)$. Note that any of the base potentials  $V_b$ in this table  can be multiplied by any of 
the convex potentials
$V_c$ to construct the Margarita potential $V(\phi) = V_b(\phi)\times V_c(\phi)$. ($m_p=1$ is assumed in this table).}
\end{minipage}
\end{center}
\end{table*}

A list of some plateau base potentials $V_b$ and convex potentials $V_c$
is given in table 1. 
Note that any of the base potentials $V_b$ in table 1 can be multiplied by any of the convex potentials
$V_c$ to construct the Margarita potential $V(\phi) = V_b(\phi)\times V_c(\phi)$. 
These examples are by no means exhaustive. However a general discussion of the Margarita potential
both in the context of inflation and dark energy lies beyond the scope of the present paper.
\footnote{In the context of the convex potential $e^{\beta\phi^2}$ one should note that the resulting
Margarita potential will satisfy equipartition initial conditions
$V(\phi) \sim {\dot\phi}^2 \sim \mpl^4$ but the extreme steepnes of $e^{\beta\phi^2}$ will prevent inflation
from occuring while the inflaton rolls down the steep branch of the Margarita potential.
Consequently the presence of a small initial (positive) curvature term could quite
easily dominate over $V(\phi)$ and make the universe contract before inflating.}

\section{Inflation in the non-canonical framework}
\label{sec:inf_noncan}

 A natural extension of  the inflationary framework involves  Lagrangian with  a non-canonical kinetic term \cite{Mukhanov:2005bu}. An attractive feature of this class of models is  that the equations of motion remain second order \cite{Unnikrishnan:2012zu,Mukhanov:2005bu}.

Non-canonical scalars have the Lagrangian density \cite{Mukhanov:2005bu}
\beq
{\cal L}(X,\phi) = X\l(\frac{X}{M^{4}}\r)^{\alpha-1} -\; V(\phi), ~~~~ X = \frac{1}{2}{\dot\phi}^2 \, ,
\label{eq:lag_nc}
\eeq
where $M$ has dimensions
 of mass, while $\alpha \geq 1$ is a dimensionless parameter. When $\alpha = 1$
the Lagrangian (\ref{eq:lag_nc}) reduces to the usual canonical scalar field Lagrangian
${\cal L}(X,\phi) = X -\; V(\phi)$.

The energy density and pressure have the form
\ber
\rho_{_{\phi}} &=& \l(2\alpha-1\r)X\l(\frac{X}{M^{4}}\r)^{\alpha-1} +\;  V(\phi) \, ,\nonumber\\
p_{_{\phi}} &=& X\l(\frac{X}{M^{4}}\r)^{\alpha-1} -\; V(\phi) \, , ~~
X \equiv \frac{1}{2} {\dot \phi}^{2} \, ,
\label{eq:rho_p_nc}
\eer
which reduces to the canonical expression
$\rho_{_{\phi}} = X + V$, ~$p_{_{\phi}} = X - V$ when $\alpha = 1$.

One should note that the
equation of motion
\beq
{\ddot \phi}+ \f{3\, H{\dot \phi}}{2\alpha -1} + \l(\f{V'(\phi)}{\alpha(2\alpha -1)}\r)\l(\f{2\,M^{4}}{{\dot \phi}^{2}}\r)^{\alpha - 1} =\; 0 
\label{eq:EOM_nc}
\eeq
is singular at $\dot{\phi} \rightarrow 0$ and needs to be
regularized so that
 the value of $\ddot{\phi}$ remains finite
in this limit. This can be done \cite{Unnikrishnan:2012zu}  by modifying the Lagrangian (\ref{eq:lag_nc})
to 
 \beq
{\cal L}_{_R}(X,\phi) = \l(\frac{X}{1 + \beta}\r)\l(1 + \beta\l(\frac{X}{M^{4}}\r)^{\alpha-1}\r) -\; V(\phi) \, ,
\label{eq:lag_nc_new}
\eeq
where $\beta$ is a dimensionless parameter.
In the limit when $\beta \gg 1$, equation (\ref{eq:EOM_nc})
 can be approximated as
\beq
{\ddot \phi}\,+\, \f{3\, H{\dot \phi}}{2\alpha -1}\, +\, \l(\f{V'(\phi)}{\epsilon\, +\, \alpha(2\alpha -1)\l(X/M^{4}\r)^{\alpha - 1}}\r)\, =\; 0, ~~ X = \frac{1}{2}{\dot \phi}^2 \, ,
\label{eq:EOM_nc_new}
\eeq
where $\epsilon \equiv (1 + \beta)^{-1}$ is an infinitesimally small correction factor  when $\beta >> 1$.

As shown in \cite{Unnikrishnan:2012zu} for potentials behaving like
$V(\phi) \propto \phi^p$ near the minimum,  the average EOS during scalar field oscillations is
\beq
\langle w_{\phi} \rangle = \frac{p\, -\, 2\alpha}{p\,(2\alpha -1)\, +\, 2\alpha} \, .
\label{eq:EOS_nc}
\eeq
For $\alpha = 1$ the above expression reduces to the canonical result 
\beq
\langle w_{\phi} \rangle = \f{p-2}{p+2}\, .
\label{eq:EOS_avg_can}
\eeq
The inflationary slow-roll parameter $\epsilon_{nc}$ for non-canonical inflation  is 
given by \cite{Unnikrishnan:2012zu}
\begin{equation}
\epsilon_{nc}=\left(\frac{1}{\alpha}\right)^{\frac{1}{2\alpha-1}}\left(\frac{3M^4}{V}\right)^{\frac{\alpha-1}{2\alpha-1}}\left(\epsilon_c\right)^{\frac{\alpha}{2\alpha-1}} \, ,
\label{eqn:slow-roll_nc}
\end{equation}
$\epsilon_c$ being the canonical slow-roll parameter (\ref{eq:CMB_eV}).
Note that $\epsilon_{nc}<\epsilon_c$ for $3M^4\ll V$. This suggests
that for a fixed potential $V$,  the duration of inflation can be enhanced relative to
the canonical case
($\alpha = 1$), by a suitable choice of $M$. 

In the following, we will  discuss the   inflationary predictions for   $\lbrace \ns,r \rbrace$ in the non-canonical framework,  starting with the family of   monomial  power law potentials $V(\phi) \propto \phi^p$, followed by the  inverse  power law potentials $V(\phi) \propto \phi^{-p}$.

\subsection{Monomial  power law potentials in the non-canonical framework}
\label{sec:non_can_powerlaw}

Monomial  power law potentials in the non-canonical framework are described by the Lagrangian density

\beq
{\cal L}(X,\phi) = X\l(\frac{X}{M^{4}}\r)^{\alpha-1} -\; V_0 \, \l(\f{\phi}{m_p}\r)^p \, .
\label{eq:lag_nc_powerlaw}
\eeq

Expressions for  the scalar spectral index $\ns$, and the tensor-to-scalar ratio $r$ are given by \cite{Unnikrishnan:2012zu}

\ber
\ns = 1 - 2 \,  \l(\f{ \gamma + p}{2\, \gamma \, N_* + p}\r) \, ,  \label{eq:ns_nc_powerlaw}\\
r = \f{1}{\sqrt{2\alpha-1}}\l(\f{16 \, p}{2 \, \gamma \, N_* + p}\r)\, ,
\label{eq:r_nc_powerlaw}
\eer

where the parameter $\gamma$ is defined as

\beq
\gamma = \f{2 \, \alpha + p(\alpha-1)}{2\, \alpha-1}\, .
\label{eq:nc_gamma}
\eeq

The tensor spectral tilt for non-canonical  potentials satisfies the single field consistency relation  \cite{Unnikrishnan:2012zu} 
\beq
r = -\f{8}{\sqrt{2\alpha-1}}\, \nt \, ,
\label{eq:nc_consis_slow-roll}
\eeq
which differs from the canonical consistency relation (\ref{eq:consis_slow-roll}) for $\alpha>1$, and hence can be used as a smoking gun test for non-canonical inflation.

\begin{figure}[htb]
\begin{center}
\includegraphics[width=0.8\textwidth]{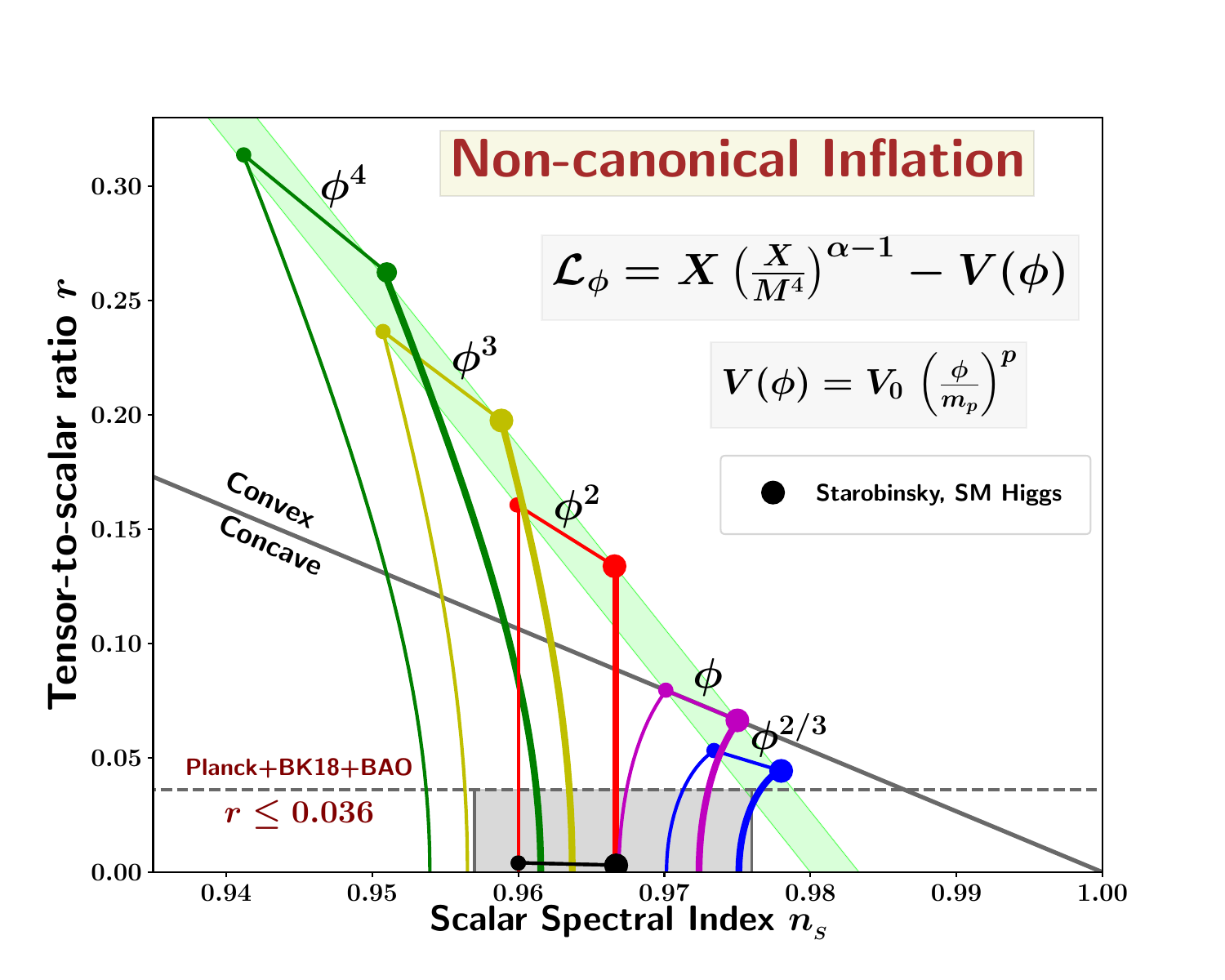}
\caption{This figure is a plot of  the  tensor-to-scalar ratio $r$, versus  the  scalar spectral index $\ns$, for the  monomial  power law potentials, $V \propto \phi^p$,  in the non-canonical framework (\ref{eq:lag_nc_powerlaw})  (the thinner and thicker   curves correspond to $N_* = 50,\, 60$ respectively). The latest CMB 2$\sigma$ bound 
$0.957 \leq \ns \leq 0.976$
 and
 the upper bound on the tensor-to-scalar ratio  $r\leq 0.036$ are
 indicated by the shaded grey colour region. The predictions of  power law potentials in the non-canonical framework appear to be somewhat similar to those of the T-model potential (\ref{eq:pot_Tmodel}) in the canonical framework (see figure \ref{fig:inf_ns_r_Tmodel}), with the parameter $\alpha$ in (\ref{eq:lag_nc_powerlaw}) playing a  role similar to that of $\lambda$ in (\ref{eq:pot_Tmodel}) (note that $r$ decreases upon increasing the value of $\alpha$ in accordance with equation (\ref{eq:r_nc_powerlaw})  and this model becomes compatible with  CMB data for large enough values of $\alpha$ as shown in table \ref{table:app4} of appendix \ref{sec:app_bounds}). However, for small values of $r$, the $r$ versus $\ns$ curves  of the non-canonical  potential , $V \propto \phi^p$,
do not converge towards a common  cosmological attractor for different values of $p$ (in contrast to the attractor behaviour of the T-model, shown in figure \ref{fig:inf_ns_r_Tmodel}). Instead, they span virtually 
the entire parameter space of the observationally allowed region in  the $\lbrace \ns,r \rbrace$ plane.}
\label{fig:inf_ns_r_nc_powerlaw}
\end{center}
\end{figure}

\begin{figure}[htb]
\begin{center}
\includegraphics[width=0.8\textwidth]{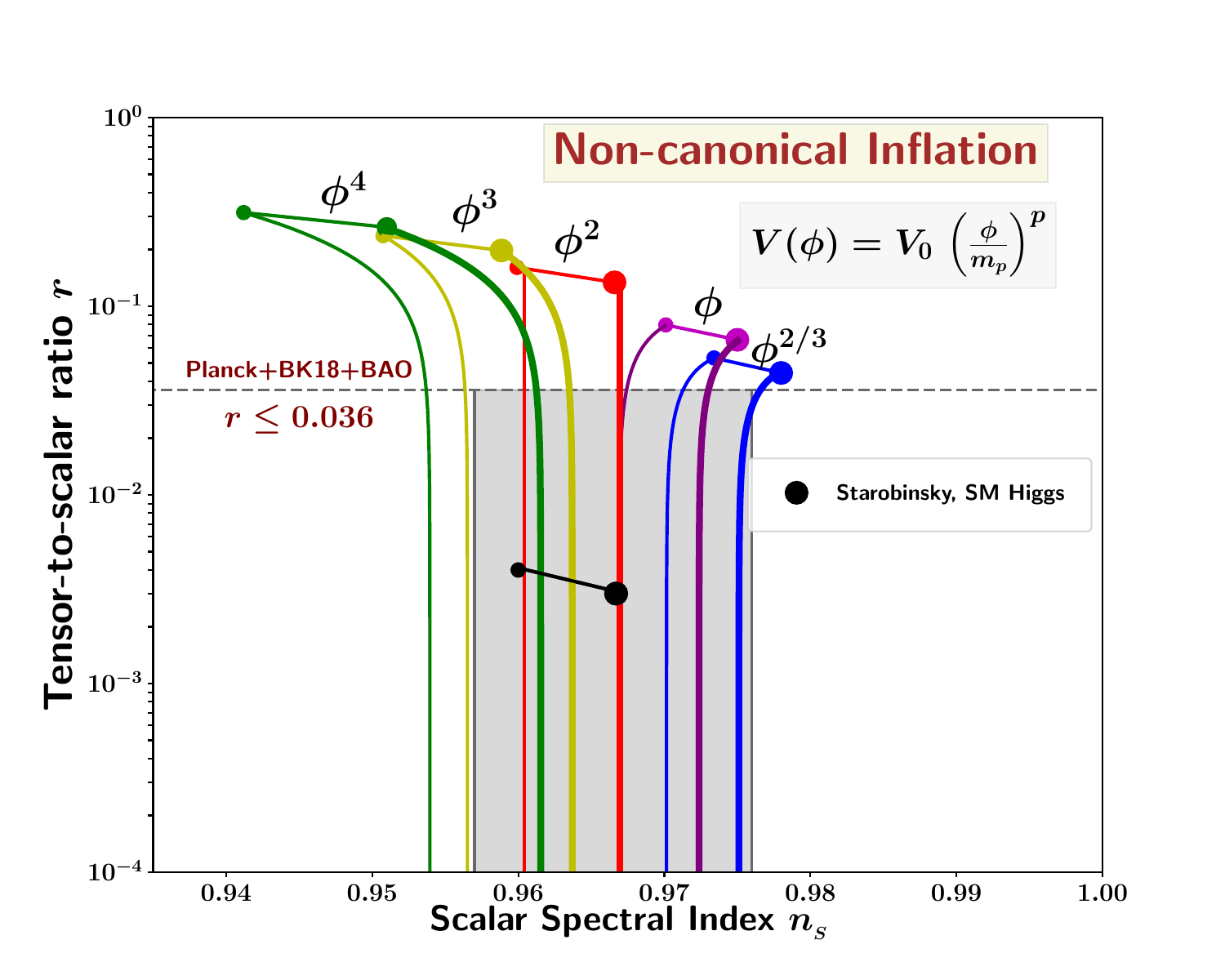}
\caption{This figure plots the  tensor-to-scalar ratio $r$, versus  the  scalar spectral index $\ns$, for monomial  power law potentials, $V \propto \phi^p$, in the non-canonical framework (\ref{eq:lag_nc_powerlaw}). The latest CMB 2$\sigma$ bound 
$0.957 \leq \ns \leq 0.976$
 and
 the upper bound on the tensor-to-scalar ratio  $r\leq 0.036$ are
 indicated by the shaded grey colour region. The tensor-to-scalar ratio $r$ in this figure has been plotted in the logarithmic scale in order to illustrate the asymptotic behaviour of $\lbrace \ns,r \rbrace$ in the limit $\alpha \gg 1$, as given in (\ref{eq:ns_r_nc_nonatt}), which shows that larger values of $\alpha$  result in smaller values of $r$. We notice  that the $\lbrace \ns,r \rbrace$ predictions in the  large $\alpha$ limit are different for different values of $p$, demonstrating the absence of cosmological attractor behaviour  which is present in
 the T-model (\ref{eq:pot_Tmodel}) as shown in figure \ref{fig:inf_ns_r_Tmodel}.}
\label{fig:inf_ns_r_nc_powerlaw_log}
\end{center}
\end{figure}

The plots of  $r$ versus $\ns$ in the non-canonical framework  for  power law potentials 
$V \propto \phi^p$  are shown in figure \ref{fig:inf_ns_r_nc_powerlaw} for  different values of $p$. For $\alpha=1$, the predictions of non-canonical  power law potentials match with that of the canonical  power law potentials, as expected. An increase in the value of $\alpha$ leads to a decrease in the tensor-to-scalar ratio $r$ for all values of $p$. Hence in the non-canonical framework,  power law potentials  can be consistent with the  CMB constraints for large enough values of $\alpha$ (see table \ref{table:app4}). In the large $\alpha$ limit, the $\lbrace \ns,r \rbrace$ predictions asymptote to

\beq
\ns = 1 - \f{3 \, p + 2}{(p+2) \, N_*  + p } \, , ~~   r = \f{1}{\sqrt{2 \alpha-1}}\l(\f{16 \, p}{(p+2) \, N_* + p}\r)\, , ~{\rm for}~ \alpha \gg 1\, ,
\label{eq:ns_r_nc_nonatt}
\eeq
which is  illustrated in figure \ref{fig:inf_ns_r_nc_powerlaw_log}. 

From figure \ref{fig:inf_ns_r_nc_powerlaw}, we notice that the predictions of  power law potentials in the non-canonical framework appear to be somewhat similar to that of the T-model potential (\ref{eq:pot_Tmodel}) in the canonical framework (see figure \ref{fig:inf_ns_r_Tmodel}), with the non-canonical  parameter $\alpha$ in (\ref{eq:lag_nc_powerlaw}) playing a  role similar to that of $\lambda$ in the T-model (\ref{eq:pot_Tmodel}). However  the $r$ versus $\ns$ plots in  figure \ref{fig:inf_ns_r_nc_powerlaw} are curved lines, in contrast to those in the case of T-model in figure \ref{fig:inf_ns_r_Tmodel} (where they are nearly straight lines). Another important difference is that the predictions of non-canonical  power law potentials for different values of $p$,  as given in (\ref{eq:ns_r_nc_nonatt}), do not converge towards a  cosmological attractor in the large $\alpha$ limit, 
in contrast to the predictions of the T-model  which approach the cosmological attractor (\ref{eq:ns_r_Tmodel_att}) independently
 of the value of $p$. The absence of cosmological attractor behaviour for non-canonical  power law potentials  implies that they are capable of scanning through the entire parameter space of the observationally allowed region in the  $\lbrace \ns,r \rbrace$ plane, as can be seen from figures \ref{fig:inf_ns_r_nc_powerlaw} and \ref{fig:inf_ns_r_nc_powerlaw_log}. Hence they are very important target candidates for future CMB missions.

\bigskip

Before proceeding to discuss inverse  power law potentials in the non-canonical framework, we would like to briefly discuss the Standard Model Higgs inflation.

\subsection{Standard Model Higgs inflation in the non-canonical framework}
\label{sec:non_can_Higgs}

The Standard Model (SM) Higgs potential is given by 
\beq
V(\phi) =  \f{\lambda}{4} \l(\phi^2 - \sigma^2\r)^2 \, ,
\label{eq:pot_SM_Higgs_actual}
\eeq
where  $\sigma$ is  the vacuum expectation value of the SM Higgs field 
\begin{equation}
\sigma=246 \, {\rm GeV} =1.1\times 10^{-16} \,  m_{p} \, ,
\label{eqn:higgs_vev}
\end{equation}  
and the Higgs  self-coupling constant has the value  $\lambda \simeq 0.1$.
It would be very interesting if 
inflation could be sourced by the SM Higgs field. 
Unfortunately, in the canonical framework, the self-interaction coupling of the Higgs field, 
$\lambda$ in (\ref{eq:pot_SM_Higgs_actual}), is far too large to be
consistent with the small amplitude of scalar fluctuations observed
by the CMB  which suggest the much smaller value $\lambda \sim 10^{-13}$
\cite{Planck_inflation}. 
However, this situation can  be remedied if either of the following two possibilties
is realized:
\begin{enumerate}
\item The Higgs couples non-minimally to gravity \cite{Fakir:1990eg,Bezrukov:2007ep,Mishra:2018dtg}. 
\item The Higgs field is described by a non-canonical Lagrangian \cite{Unnikrishnan:2012zu,Mishra:2018dtg}.
\end{enumerate}

In section \ref{sec:plateau_can}, we briefly discussed the first possibility and the corresponding predictions of $\lbrace \ns,r \rbrace$. Here we focus on SM Higgs inflation in the non-canonical framework.
\bigskip

Given that $\sigma = 246\, {\rm GeV} \ll m_p$, the Higgs potential (\ref{eq:pot_SM_Higgs_actual}) in the limit $\phi \gg \sigma$ takes the form 

\beq
V(\phi) \simeq  \f{\lambda}{4} \phi^4 \, ,
\label{eq:pot_SM_Higgs_approx}
\eeq
which is the quartic monomial  power law potential.  In the non-canonical framework, the $\lbrace \ns,r \rbrace$ predictions  of SM Higgs potential becomes 
\ber
n_{_{S}}= 1 -\l(\f{\gamma + 4}{N_*\gamma + 2}\r) \, , \\
\label{eqn:n_s}
r =  \l(\f{1}{\sqrt{2\,\alpha - 1}}\r)\l(\f{32}{N_*\gamma + 2}\r) \, ,
\label{eqn: T-to-S phi-n pot}
\eer
where 
\ber
\gamma \equiv \frac{2(3\alpha - 2)}{2\alpha - 1} \, .
\label{eqn: gamma}
\eer
Since $\gamma$ increases from $\gamma = 2$ for $\alpha = 1$ to
$\gamma = 3$ for $\alpha  \gg 1$, therefore
the scalar spectral index  increases
from the canonical value $n_{_{S}} = 0.951$ ($\alpha = 1, N_* = 60$) to $n_{_{S}} = 0.962$, 
in non-canonical models (with $\alpha  \gg 1$). Similarly  the tensor-to-scalar ratio $r$ declines with the increase in $\alpha$ as observed before in figure \ref{fig:inf_ns_r_nc_powerlaw}. Note that the black colour dots in  figure  \ref{fig:inf_ns_r_nc_powerlaw} represent the $\lbrace \ns,r \rbrace$ predictions of the non-minimally coupled SM Higgs inflation, while the green curves represent the same for SM Higgs inflation in the non-canonical framework.
 
 \begin{figure}[htb]
\begin{center}
\includegraphics[width=0.8\textwidth]{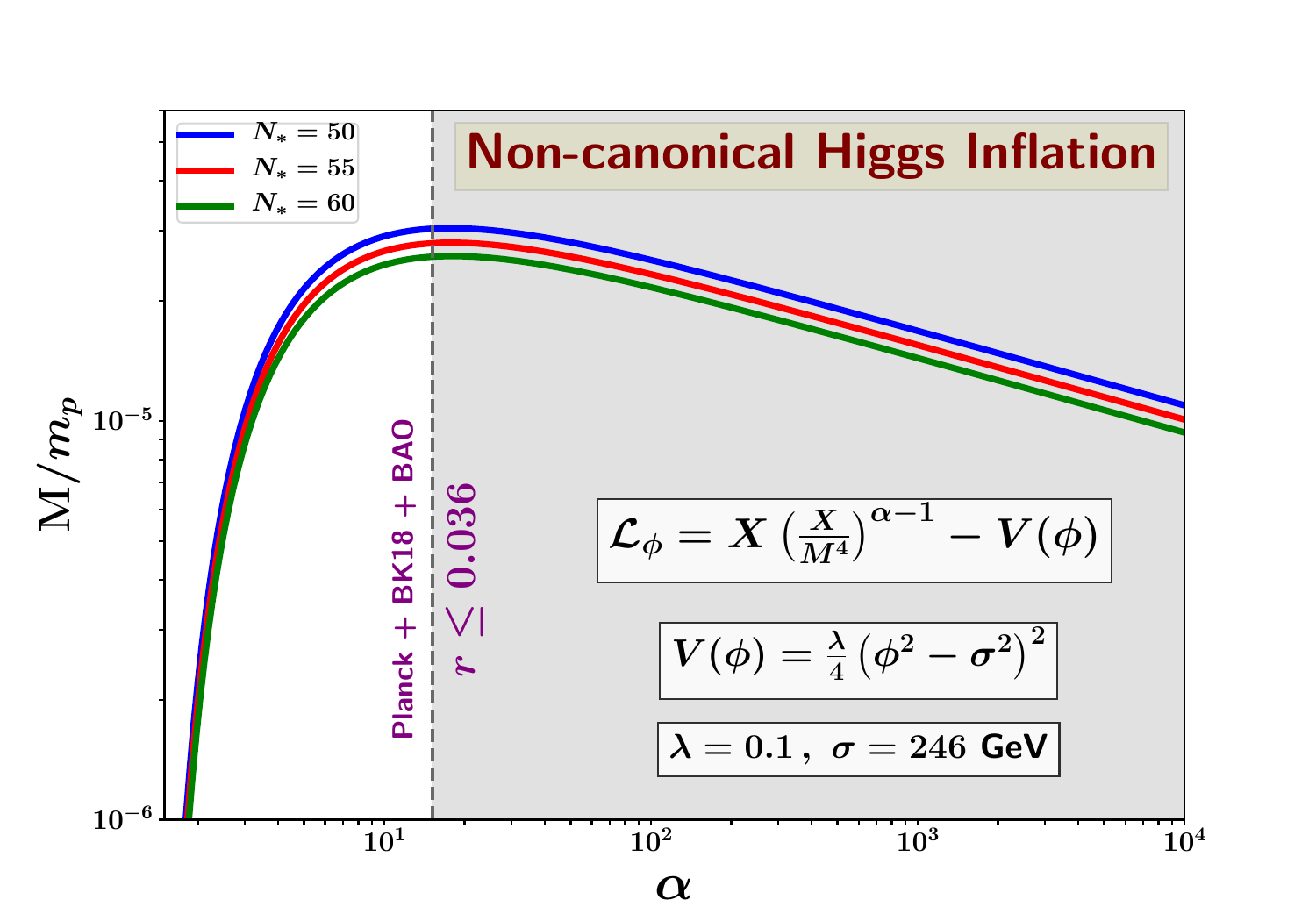}
\caption{This figure illustrates the relation between the non-canonical parameters $M$ and $\alpha$, given by equation (\ref{eqn:lambda_vs_lambdac}), for SM Higgs inflation which results in the 
self-coupling value $\lambda=0.1$ in equation (\ref{eq:pot_SM_Higgs_actual}).
 Results for three different e-folding values $N_*=50,~55,~60$ are shown by blue, red, and green colour curves respectively. The shaded region corresponds to observationally allowed values of  $\alpha$ obeying the CMB constraint $r\leq 0.036$.}
\label{fig:inf_nc_Malpha}
\end{center}
\end{figure}
 
 The relation between the value of the Higgs self-coupling $\lambda\simeq 0.1$ in the non-canonical 
framework and the corresponding canonical value $\lambda_c$ is given by \cite{Mishra:2018dtg}
\ber
\lambda = 4\left[\frac{32\lambda_c(N_*+1)^3}{\sqrt{2\alpha-1}}\left(\frac{\alpha}{4}\Big(\frac{1}{6}\frac{m_p^4}{M^4}\Big)^{\alpha-1}\right)^{\frac{2}{3\alpha-2}}\left(\frac{1}{N_*\gamma+2}\right)^{\frac{\gamma+4}{\gamma}}\right]^{\frac{3\alpha-2}{\alpha}} \, ,
\label{eqn:lambda_vs_lambdac}
\eer
where consistency with CMB observations suggests $\lambda_c \sim 10^{-13}$.

 Figure \ref{fig:inf_nc_Malpha}  describes the values of the non-canonical parameters $\alpha$ and $M$ that yield $\lambda\simeq 0.1$
 in (\ref{eq:pot_SM_Higgs_approx}) 
-- the relation between $M$ and $\alpha$ being provided by
 equation (\ref{eqn:lambda_vs_lambdac}). From our analysis, it is clear that SM Higgs field  in the non-canonical framework can successfully source inflation, while obeying the CMB constraints, for large enough values of the non-canonical parameter $\alpha$, as shown by the shaded region in figure \ref{fig:inf_nc_Malpha}.

\section{Discussion}
\label{sec:discussion}

 The latest CMB  results from the BICEP/Keck collaboration \cite{BICEP:2021xfz}, combined with the PLANCK 2018 data release \cite{Planck_inflation}, have imposed a stringent upper bound on the primordial tensor-to-scalar ratio $r\leq 0.036$ (at $95\%$ confidence).  Additionally,  CMB observations constrain the scalar spectral tilt to be $n_{_S} \in [0.957,0.976]$ at the 2$\sigma$ level. In this paper, we have discussed the implications of these latest CMB constraints on inflationary models, with particular emphasis on inflation sourced by a scalar field with a  non-canonical Lagrangian density.
 
  In section \ref{sec:inf_can}, we discussed the implications of the latest CMB constraints  on the  primordial observables  $\lbrace \ns,r\rbrace$  in the  framework of  slow-roll inflation driven by a  single canonical scalar field $\phi$ with a potential $V(\phi)$. The upper bound on the tensor-to-scalar ratio $r\leq 0.036$ strongly disfavours monotonically increasing convex 
inflationary potentials within the canonical framework. In fact, one of the central inferences that can be  drawn from the CMB constraints is the fact that  the entire family of monomial power law potentials $V(\phi) \sim \phi^p$ are now completely ruled out in the canonical framework. Among these, are the simplest classic inflationary models including the $\f{1}{2}m^2\phi^2$ and  $\lambda\phi^4$ potentials.  Instead, the CMB data now favours asymptotically flat plateau like potentials.  In section \ref{sec:plateau_can}, we discussed a number of important plateau potentials in light of the latest observational constraints, including the Einstein frame potentials associated with Starobinsky inflation and  non-minimally coupled  Higgs inflation, as well as  the T-model and E-model $\alpha$-attractors and the D-brane KKLT inflation. The $\lbrace \ns,r\rbrace$ predictions of these models are illustrated in figure \ref{fig:inf_ns_r_latest}.  In the limit $r\ll 1$, the  $\lbrace \ns,r\rbrace$ predictions of  $\alpha$-attractor potentials exhibit  an attractor behaviour \cite{Kallosh:2019hzo,Kallosh:2021mnu} as illustrated in figure \ref{fig:inf_ns_r_Tmodel} for the T-model.  We demonstrated that plateau potentials generically produce  
a small tensor-to-scalar ratio and hence are consistent with the CMB data, in agreement with earlier results 
\cite{Kallosh:2019hzo,Kallosh:2021mnu}. 
  
       However, for large field values, the plateau potentials saturate to a constant value $V_0$ which is typically small, namely $V_0 \leq 10^{-10}\, m_p^4$ for $r\ll 1$. The small height of a plateau potential bears significant implications
for the issue of initial conditions for inflation. Owing to the fact that the plateau height is typically very small compared to the Planck scale,  an  equipartition of the initial  kinetic, potential, and gradient energy terms  is not possible at very early times when initial conditions for inflation are usually set. Similarly, the initial presence of positive spatial curvature 
could also be problematic for inflation with plateau potentials. Additionally, for $r\ll1$, plateau potentials might exhibit 
a large running of the scalar spectral tilt $\ns$ as recently pointed out in \cite{Easther:2021rdg}. Due to the aforementioned reasons, we focus on   scenarios where monotonically increasing convex potentials   can be accommodated by the CMB data. In this 
connection, we study two types of models. The first one is the Margarita potential formulated in the canonical framework, 
while the second scenario  is associated with  non-canonical Lagrangians of the form ${\cal L} = X^\alpha - V(\phi)$,
\, $\alpha \geq 1$ ($X = {\dot\phi}^2/2$). 
       
       In section \ref{sec:margarita}, we introduced the Margarita potential (\ref{eq:pot_margarita_gen}) which features a monotonically growing convex   inflationary wing  at the commencement of  a  plateau potential. 
We discussed the $\lbrace \ns,r \rbrace$ predictions of a
specific Margarita potential (\ref{eq:pot_margarita_T_exp}) constructed by incorporating 
extensions to the T-model potential at large field values. We demonstrated that Margarita potentials easily
satisfy CMB constraints and also allow for an   equipartition between  initial energy densities. We also discussed a general recipe for constructing Margarita potentials. 
       
       Section \ref{sec:inf_noncan} was dedicated to the study of inflation in the non-canonical 
framework \cite{Unnikrishnan:2012zu,Unnikrishnan:2013vga} in the light of the recent CMB data release. 
In section \ref{sec:non_can_powerlaw}, working with  the family of power law potentials $V(\phi) \sim \phi^p$, we demonstrated that inflation with monomial power law potentials can be successfully resurrected  in the non-canonical framework. In particular, we compared and contrasted the $\lbrace \ns,r\rbrace$ predictions of non-canonical power law potentials shown in figure \ref{fig:inf_ns_r_nc_powerlaw} against those of the T-model $\alpha$-attractor potential in the canonical framework shown in figure \ref{fig:inf_ns_r_Tmodel}. We found that CMB predictions of power law potentials  in non-canonical inflation do not exhibit attractor 
behaviour, which stands in contrast to the predictions of $\alpha$-attractors. Instead, they span the entire parameter space of the observationally allowed region in  the $\lbrace \ns,r \rbrace$ plane.    This is one of the central results of our analysis.  In section \ref{sec:non_can_Higgs}, we discussed the possibility of realising successful inflation with the Standard Model Higgs potential in the non-canonical framework. 
We showed that the npn-canonical Higgs with standard model parameters could easily satisfy CMB constraints.
       
       Section \ref{sec:non_can_IPL} was devoted to a study of inflation with  inverse power law (IPL) potentials 
$V(\phi) \sim \phi^{-p}$ in the non-canonical framework. IPL potentials within the non-canonical framework
 lead to power law inflation with $a(t)\sim t^q$ \cite{Unnikrishnan:2013vga}. We demonstrated
  that the $\lbrace \ns,r \rbrace$ predictions of   IPL potentials are consistent with the CMB data for sufficiently 
large values of the non-canonical parameter $\alpha > 20$,  as shown in figure \ref{fig:inf_ns_r_nc_IPL}.  
We also discussed a remedy to  the graceful exit problem for power law inflation in the non-canonical framework. 

\bigskip

Before concluding,  we  succinctly summarise  the analysis carried out in this paper for  the sake of clarity.

\begin{itemize}

\item Given the latest CMB constraints  on $\lbrace n_{_S}, r\rbrace$, obtain explicit bounds on important inflationary parameters such as the Hubble slow-roll parameters $\epsilon_H, \, \eta_H$, the potential slow-roll parameters $\epsilon_{_V}, \, \eta_{_V}$,  the Hubble parameter $H_k^{\rm inf}$, equation of state of the universe during inflation $w_\phi$, energy scale of inflation $E^{\rm inf}$ and the tensor tilt $n_{_T}$ in the context of single field slow-roll inflation. 

\item  We highlight  that the latest upper bound on the tensor-to-scalar ratio, namely $r \leq 0.036$, not only rules out   the classic convex potentials $V(\phi) \sim \phi^2,\, \phi^4$, but also disfavours the entire family of monomial potentials $V(\phi) \sim \phi^p$ for any $p>0$ in the canonical framework (at $95\%$ confidence). While this result was highlighted  in \cite{BICEP:2021xfz,Kallosh:2021mnu}, we discuss it in section \ref{sec:inf_can} in order to make comparison with the predictions of non-canonical monomial potentials in section \ref{sec:inf_noncan}.

\item We stress that the latest CMB data favours asymptotically flat potentials\footnote{ This result is already well-known in the inflationary literature, however, we discuss them for the continuity of the flow of the discussions in this paper}  in the canonical framework and discuss a few examples on asymptotically flat potentials  that belong to either of the list of categories - symmetric/asymmetric, one/two parameter models, exponential/algebraic approach to plateau behaviour.

\item We then critically scrutinize the problem of initial conditions for these plateau potentials and propose the Margarita potential (which is, in fact, a class of potentials) that can  address the issue of initial conditions. We also discuss a generic recipe to construct the  Margarita potential phenomenologically.  

\item We then focus on inflation in the non-canonical framework and demonstrate that convex monomial potentials, that are disfavoured in the canonical framework, become compatible with the CMB data in the con-canonical framework for a suitable range of values of the non-canonical parameter $\alpha$.  Similarly, in appendix \ref{sec:non_can_IPL} we show that the inverse  power law  potential $V(\phi) \sim \phi^{-p}$, which leads to  power law inflation in the non-canonical framework,  also satisfies the latest CMB bounds.  While some of these results were earlier discussed in \cite{Unnikrishnan:2012zu,Unnikrishnan:2013vga}, we reproduce the results and discuss their implications  in light of the recent data. 

\item During our analysis in the non-canonical  framework, we discover striking similarities between the $\lbrace n_{_S}, r\rbrace$ flow lines of  monomial potentials $V(\phi) \sim \phi^p$ in the non-canonical framework and the  T-model $\alpha$-attractors canonical framework. We also discuss the difference between the predictions of both the models. This is one of the highlights of our analysis. 

\end{itemize}

\bigskip
       
       Finally we would like to draw attention to the fact
 that the analysis of initial conditions for inflation with plateau potentials \cite{Mishra:2018dtg} remains an interesting open area of research, especially in presence of a positive spatial curvature \cite{Linde:2014nna,Guth:2013sya}. 
We shall study this in more detail in a companion  work \cite{s_v_t_curvature}. 
However, we would  also like to emphasise that regardless of the issue of initial conditions, both 
power law and inverse power law potentials  in the non-canonical framework (and the Margarita potential in the canonical
framework), satisfy current CMB constraints and are therefore important target  candidates for the  next generation
of CMB missions.

\section{Acknowledgements}
SSM is supported by a STFC Consolidated Grant [Grant No. ST/T000732/1].  VS was partially supported by the J.~C.~Bose Fellowship of Department of Science and Technology, Government of India. SSM thanks IUCAA (India) for their hospitality.

\medskip

For the purpose of open access, the authors have applied a CC BY public copyright license to any Author Accepted Manuscript version arising. \\

\medskip

{\bf Data Availability Statement:} This work is entirely theoretical and has no associated data.

\appendix

\section{Constraint  on parameters of inflationary potentials }
\label{sec:app_bounds}

In this appendix, we present bound on the  parameters of different inflationary potentials discussed in the paper, namely, $\lambda$ in T-model and E-model $\alpha$-attractors, $M$ in D-brane KKLT potential and the non-canonical parameter $\alpha$ in the context of non-canonical inflation with  monomial potential $V(\phi)\sim \phi^p$. We obtain these constraints numerically by imposing the CMB $2\sigma$ bound on $\ns$ and $r$ given in (\ref{eq:CMB_ns_obs}) and (\ref{eq:CMB_r_obs}) -

$$
n_{_S} \in [0.957,0.976]~, ~~~r \leq 0.036~.
$$

\begin{table}[htb]
\begin{center}
 \begin{tabular}{|c|c|c|c|c|c|c|}
 \hline
 \Tstrut

 $V(\phi) =V_0 \, \tanh^{p}{\l(\f{\lambda \phi}{m_p}\r)} $ & $N_* = 50$   &  $N_* = 60$  \\ [1ex]
  \hline\hline \Tstrut
  $p = 1$ & $\lambda \geq 0.112$ & $\lambda \geq 0.086$  \\ [1.2ex] 
 \hline  \Tstrut
 $p = 2$ & $\lambda \geq 0.132$ & $\lambda \geq 0.108$  \\ [1.2ex] 
  \hline  \Tstrut
 $p = 3$ & $\lambda \geq 0.138$ & $\lambda \geq 0.113$  \\ [1.2ex] 
 \hline  \Tstrut
 $p = 4$ & $\lambda \geq 0.141$ & $\lambda \geq 0.116$  \\ [1.2ex] 
 \hline
\end{tabular}
\captionsetup{
	justification=raggedright,
	singlelinecheck=false
}
\caption{CMB $2\sigma$ bound on the  parameter $\lambda$ of   T-model $\alpha$-attractor potential (\ref{eq:pot_Tmodel}) is presented in this table for different values of $p$. }
\label{table:app1}
\end{center} 
\end{table}

\begin{itemize}
\item We begin with T-model $\alpha$-attractor  potential  (\ref{eq:pot_Tmodel})

$$V(\phi)  =V_0 \, \tanh^{p}{\l(\lambda \f{\phi}{m_p}\r)} ~,$$
and present constraints on the parameter $\lambda$ for different values of $p$ in table \ref{table:app1}.

\item  For  E-model $\alpha$-attractor  potential   (\ref{eq:pot_Emodel})

$$V(\phi) =  V_0 \, \l( 1 - e^{-\lambda\f{\phi}{m_p}} \r)^{p}~,$$
and present constraints on the parameter $\lambda$ for different values of $p$ in table \ref{table:app2}.

\begin{table}[htb]
\begin{center}
 \begin{tabular}{|c|c|c|c|c|c|c|}
 \hline
 \Tstrut

 $V(\phi) =  V_0 \, \l( 1 - e^{- \f{\lambda \phi}{m_p}} \r)^{p}  $ & $N_* = 50$   &  $N_* = 60$  \\ [1.5ex]
  \hline\hline \Tstrut
  $p = 1$ & $\lambda \geq 0.137$ & $\lambda \geq 0.094$  \\ [1.2ex] 
 \hline  \Tstrut
 $p = 2$ & $\lambda \geq 0.194$ & $\lambda \geq 0.158$  \\ [1.2ex] 
  \hline  \Tstrut
 $p = 3$ & $\lambda \geq 0.229$ & $\lambda \geq 0.182$  \\ [1.2ex] 
  \hline  \Tstrut
 $p = 4$ & $\lambda \geq 0.243$ & $\lambda \geq 0.195$  \\ [1.2ex] 
 \hline
\end{tabular}
\captionsetup{
	justification=raggedright,
	singlelinecheck=false
}
\caption{ CMB $2\sigma$ bound on the parameter  $\lambda$ of  E-model $\alpha$-attractor potential  (\ref{eq:pot_Emodel}) is presented in this table for different values of $p$. }
\label{table:app2}
\end{center} 
\end{table}

\item We then present the constraints on the parameter $M$ of D-brane KKLT potential (\ref{eq:pot_KKLT})

$$V(\phi) = V_0 \, \f{\phi^n}{\phi^n+M^n}\, ,$$ 
for different values of $n$ in table \ref{table:app3}.

\begin{table}[htb]
\begin{center}
 \begin{tabular}{|c|c|c|c|c|c|c|}
 \hline
 \Tstrut

 $V(\phi) = V_0 \, \f{\phi^n}{\phi^n+M^n} $ & $N_* = 50$   &  $N_* = 60$  \\ [1ex]
  \hline\hline \Tstrut
  $n = 2$ & $M \leq 7.07$ & $M \leq 9.16$  \\ [1.2ex] 
 \hline  \Tstrut
 $n = 4$ & $M \leq 10.17$ & $M \leq 12.21$  \\ [1.2ex] 
  \hline  \Tstrut
 $n = 6$ & $M \leq 13.51$ & $M \leq 15.82$  \\ [1.2ex] 
 \hline
\end{tabular}
\captionsetup{
	justification=raggedright,
	singlelinecheck=false
}
\caption{CMB $2\sigma$ bound on the  parameter  $M$ of  D-brane KKLT inflation potential (\ref{eq:pot_KKLT}) is presented in this table for different values of $n$. }
\label{table:app3}
\end{center} 
\end{table}

\item Finally, we present constraints on the non-canonical parameter $\alpha$ in the context of inflation in the non-canonical framework (\ref{eq:lag_nc_powerlaw}) with a monomial potential $V(\phi)\sim \phi^p$ for different values of $p$ in table \ref{table:app4}. Note that for $p=4$, the predictions  for $\ns$  of  non-canonical monomial  inflation  is not compatible with CMB data for $N_* = 50$ (while it is compatible for $N_* = 60$) as can be seen from figure \ref{fig:inf_ns_r_nc_powerlaw}. We find that monomial potential with $p=4$ becomes compatible with CMB $\ns$ bound for $N_* \geq 54$ and the constraint on  tensor-to-scalar ratio, namely, $r \leq 0.036$   translates to $\alpha \geq 15.52$ (for $N_* = 54$) . 

\begin{table}[htb]
\begin{center}
 \begin{tabular}{|c|c|c|c|c|c|c|}
 \hline
 \Tstrut

 $V(\phi) = V_0 \, \l( \f{\phi}{m_p} \r)^p$ & $N_* = 50$   &  $N_* = 60$  \\ [1ex]
  \hline\hline \Tstrut
  $p = 1$ & $\alpha \geq 4.49$ & $\alpha \geq 3.17$  \\ [1.2ex] 
 \hline  \Tstrut
 $p = 2$ & $\alpha \geq 10.18$ & $\alpha \geq 7.24$  \\ [1.2ex] 
  \hline  \Tstrut
 $p = 3$ & $\alpha \geq 14.58$ & $\alpha \geq 10.37$  \\ [1.2ex] 
  \hline  \Tstrut
 $p = 4$ &  $--$  & $\alpha \geq 12.75$  \\ [1.2ex] 
 \hline
\end{tabular}
\captionsetup{
	justification=raggedright,
	singlelinecheck=false
}
\caption{ CMB $2\sigma$ bound on the  parameter  $\alpha$ for  non-canonical inflation with monomial  potential (\ref{eq:lag_nc_powerlaw})  is presented in this table for different values of $p$.}
\label{table:app4}
\end{center} 
\end{table}

\end{itemize}

\section{Inverse  power law potentials in the non-canonical framework}
\label{sec:non_can_IPL}

    It is well known that  within the canonical framework a spatially flat universe can expand as a power law with 
$a(t)\sim t^q$,
 if inflation is sourced by  an exponential potential \cite{Lucchin:1984yf}. However, owing to the fact that the tensor-to-scalar ratio $r$ in such models turns out to be  much larger than the upper bound set by the CMB observations, namely $r\leq 0.036$, the hope of realising  power law 
inflation is dashed in the canonical framework. Instead, CMB constraints  indicate that cosmic expansion is 
near-exponential if inflation is driven by a  canonical scalar field, as discussed in section \ref{sec:Inf_CMB_obs}.
    
    Nevertheless,  power law expansion of the form $a(t)\sim t^q$ with $q>1$, can also be realised  in the non-canonical framework (\ref{eq:lag_nc}) with an inverse  power law(IPL) potential \cite{Unnikrishnan:2013vga}
\beq
V(\phi) = V_0 \l(\f{\phi}{m_p}\r)^{-p} \, ,
\label{eq:pot_IPL}
\eeq     
where $p$ is related to the non-canonical parameter $\alpha$ by \cite{Unnikrishnan:2013vga}
\beq
p =  \f{2 \, \alpha}{\alpha - 1}\, ,
\label{eq:NC_IPL_p_alpha}
\eeq    
and the expansion rate, $a(t)\sim t^q$,  is related to the EOS of the scalar field by the usual relation\footnote{The exponent  of   power law expansion  $q$ is independent of  the value of $p$. Rather, the value of $q$ is only dependent upon the amplitude $V_0$ of the IPL potential (\ref{eq:pot_IPL}), in the sense that for a given value of $\alpha$, a different value of $V_0$ leads to a different value of $q$. Alternatively, one can keep $q$ fixed by varying both $\alpha$ and $V_0$ simultaneously (see \cite{Unnikrishnan:2013vga}).} 
\beq
q = \f{2}{3} \, \l(\f{1}{1+w_{\phi}}\r) \, .
\label{eq:NC_IPL_q_EOS}
\eeq   

Expressions for  the scalar spectral index $\ns$, and the tensor-to-scalar ratio $r$ in the non-canonical framework
are given by \cite{Unnikrishnan:2013vga}

\begin{figure}[htb]
\begin{center}
\includegraphics[width=0.8\textwidth]{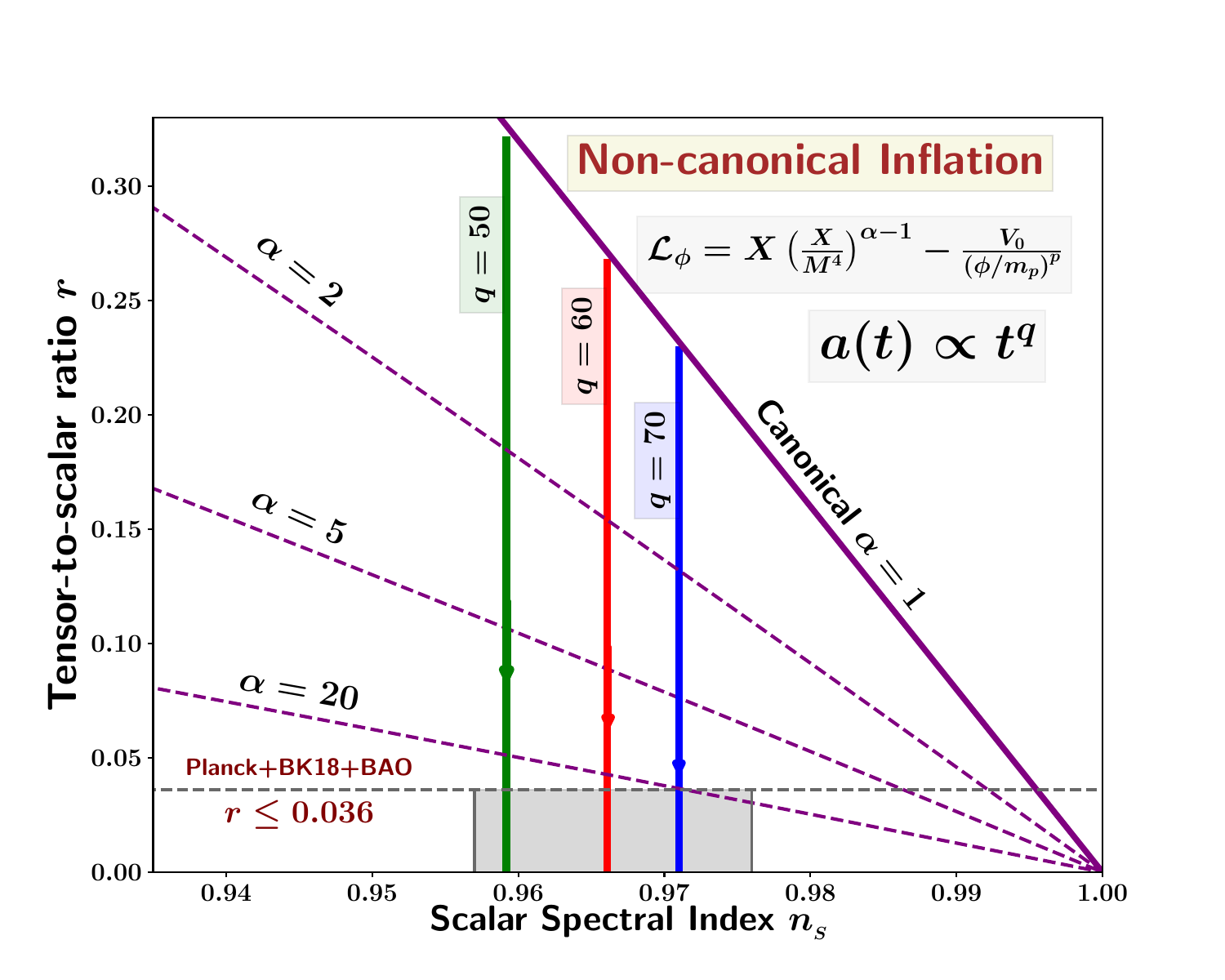}
\caption{This figure is a plot of  the  tensor-to-scalar ratio $r$, versus  the  scalar spectral index $\ns$, for the  inverse  power law (IPL) potential  (\ref{eq:pot_IPL}) in the non-canonical framework, which leads to  power law inflation $a(t)\sim t^q$, $q\gg1$. The latest CMB 2$\sigma$ bound 
$0.957 \leq \ns \leq 0.976$
 and
 the upper bound on the tensor-to-scalar ratio  $r\leq 0.036$ are
 indicated by the shaded grey colour region. The green, red, and blue  curves correspond to $q=50\, , 60\, , 70$ respectively. For a given value of $q$, while $\ns$ is insensitive to the value of the non-canonical parameter $\alpha$, the  tensor-to-scalar ratio $r$ decreases with an increase in $\alpha$, which has been illustrated by the arrow marks on each plot. The purple lines represent contours of fixed  $\alpha$ values. For $\alpha>20$ and $q\in[47.5,\, 84.3]$, predictions of 
the non-canonical  IPL potential are consistent with CMB constraints. }
\label{fig:inf_ns_r_nc_IPL}
\end{center}
\end{figure}

\ber
\ns = 1 - \f{2}{q-1} \, ,  \label{eq:ns_nc_IPL}\\
r \simeq \f{16}{q\, \sqrt{2\, \alpha-1}} \, ,
\label{eq:r_nc_IPL}
\eer
where the `$\simeq$' symbol in the expression of $r$ refers to the fact that the above equation is valid in the slow-roll limit $q\gg 1$. From  expressions (\ref{eq:ns_nc_IPL}) and (\ref{eq:r_nc_IPL}), we notice that the scalar spectral tilt  $\ns$ 
{\underline{does not depend}} upon the non-canonical parameter $\alpha$, while the tensor-to-scalar ratio $r$ decreases with an increase in $\alpha$. Hence for large enough values of $\alpha$, the IPL potential can satisfy the CMB constraints in the non-canonical framework, as shown in figure \ref{fig:inf_ns_r_nc_IPL}.

From expression (\ref{eq:ns_nc_IPL}), the CMB constraints (\ref{eq:CMB_ns_obs}) on the scalar spectral index $\ns$ translate 
into $q\in [47.5,\, 84.3]$. Figure \ref{fig:inf_ns_r_nc_IPL} depicts the behaviour of  $r$ versus $\ns$ for three different values of expansion exponent,  namely $q=50\, , 60\, , 70$, plotted in  green, red, and blue  curves respectively. The arrow mark on each plot indicates the direction of increase in the value of $\alpha$. From this figure, it is easy to see that the CMB constraints, shown by the grey colour shaded region, can easily be satisfied for $\alpha>20$. Hence  power law inflation can be successfully sourced by the IPL potential in the non-canonical framework, without violating the CMB bounds on $\lbrace \ns,r \rbrace$.

\subsection{Graceful exit from  power law inflation}

\begin{figure}[htb]
\begin{center}
\includegraphics[width=0.7\textwidth]{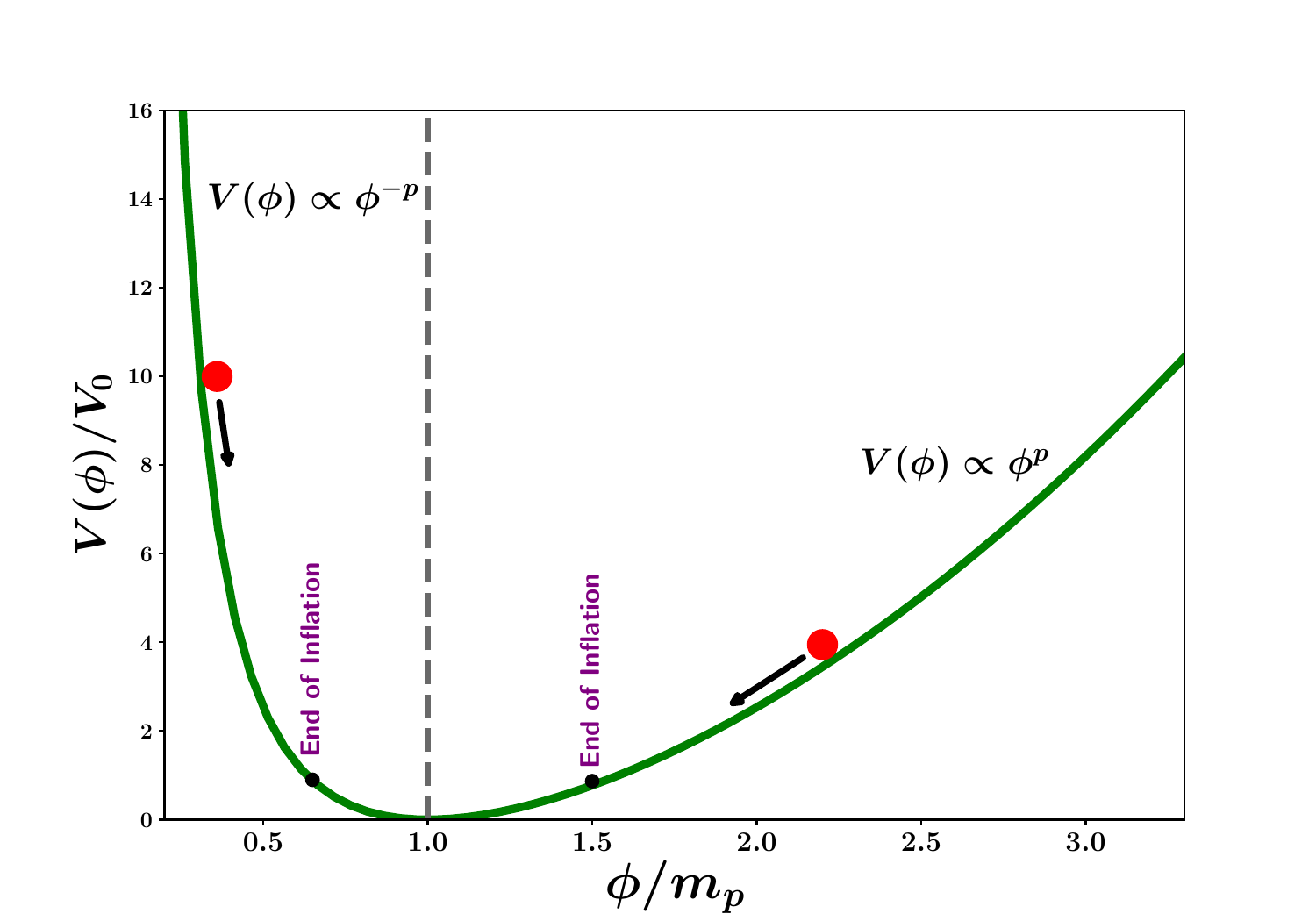}
\caption{This figure schematically depicts the  inflationary potential (\ref{eq:pot_nc_IPL_mod}) which can address the graceful exit problem of  power law inflation in the non-canonical framework. For $\phi<m_p$, the potential exhibits inverse power 
law behaviour  $V(\phi) \sim \phi^{-p}$ which results in power law inflation in the non-canonical framework. For $\phi>m_p$, the potential exhibits monomial behaviour $V(\phi) \sim \phi^{p}$ which results in quasi-exponential inflation. After the end of inflation, the non-canonical scalar field  oscillates around the  minimum of the potential at  $\phi=m_p$, resulting in successful reheating of the universe. } 
\label{fig:pot_nc_IPL_mod}
\end{center}
\end{figure}

During  power law inflation, since the universe accelerates forever, it does not account for the end of inflation and the subsequent transition into the  radiative hot Big Bang phase.  This issue of  `graceful exit'  from the inflationary phase  turns out to be one  of the central drawbacks of  power law inflation. A possible way out  is to assume that the potential driving  power law inflation  approximates a more general functional form which allows for the oscillations of the inflaton field at the end of inflation and hence results in a successful reheating scenario. An interesting general form of the potential
in the context of non-canonical inflation is \cite{Unnikrishnan:2013vga} 

\beq
V(\phi) = V_0 \l[ \l(\f{\phi}{m_p}\r)^{p/2} - \l(\f{\phi}{m_p}\r)^{-p/2} \r]^2\, ,
\label{eq:pot_nc_IPL_mod}
\eeq
with $p = 2\alpha/(\alpha-1)$ as given in (\ref{eq:NC_IPL_p_alpha}). The potential is schematically shown in figure \ref{fig:pot_nc_IPL_mod}. 

For $\phi<m_p$, the potential has  the asymptotic form  $V(\phi) \sim \phi^{-p}$ which leads to  power law inflation in the non-canonical framework. While for $\phi>m_p$, the potential exhibits monomial behaviour $V(\phi) \sim \phi^{p}$ which leads to quasi-exponential inflation. After the end of inflation, the non-canonical scalar field  oscillates around the  minimum of the potential which is located at $\phi=m_p$, resulting in successful reheating.

\printbibliography

\end{document}